\documentclass[sigplan,10pt]{acmart}
\renewcommand\footnotetextcopyrightpermission[1]{}

\usepackage{graphicx}
\usepackage{subcaption}
\usepackage{tikz}
\usepackage{pifont}
\usepackage{xspace}
\usepackage{booktabs}
\usepackage{float}
\usepackage{listings}
\usepackage{lstautogobble} 
\definecolor{codegreen}{rgb}{0,0.6,0}
\definecolor{codegray}{rgb}{0.5,0.5,0.5}
\definecolor{codepurple}{rgb}{0.58,0,0.82}
\definecolor{backcolour}{rgb}{0.95,0.95,0.92}
\definecolor{backcolour2}{rgb}{0.95,0.90,0.90}
\definecolor{lightGrey}{rgb}{0.9, 0.9, 0.9}
\definecolor{beaublue}{rgb}{0.74, 0.83, 0.9}
\definecolor{lightred}{RGB}{229, 220, 220}
\definecolor{patchparam}{RGB}{42,96,153} %

\newcommand{\Hilightt}{\makebox[0pt][l]{\color{lightred}\rule[-0.35em]{\linewidth}{1.2em}}}

\lstdefinestyle{mystyle}{
   basicstyle=\ttfamily,
   backgroundcolor=\color{backcolour},   
   commentstyle=\color{codegreen},
   numberstyle=\tiny\color{codegray},
   stringstyle=\color{codepurple},
   breaklines,
   tabsize=2,
   numbers=left,
   columns=fullflexible,
   keepspaces=true,
   frame=lines,
   numbersep=2pt,
   escapechar=|,
   captionpos=b,
   language=c,
   keywordstyle=\color{magenta},
   morekeywords={name, type, dst},
   keywordstyle=[2]\color{blue},
   morekeywords=[2]{kernel,cudaMemAdvise,cudaMallocManaged,cudaMemPrefetchAsync},
   showspaces=false,
   showstringspaces=false,
}
\newcommand{\sys}{\textsc{Ember}\xspace}
\DeclareRobustCommand{\stepc}[1]{\ding{\numexpr171+#1\relax}}
\DeclareRobustCommand{\stepn}[1]{\ding{\numexpr181+#1\relax}} %

\usepackage{ifthen}
\usepackage{mdframed}
\usepackage{xcolor}

\definecolor{rqbg}{RGB}{232,240,251}    %
\definecolor{ideabg}{RGB}{233,245,235}  %
\definecolor{sumbg}{RGB}{252,243,232}   %

\definecolor{rqrule}{RGB}{ 70,120,190}
\definecolor{idearule}{RGB}{ 80,150, 95}
\definecolor{sumrule}{RGB}{205,140, 50}

\newcommand{\MP}[1]{\textcolor{red}{#1}}

\newboolean{publicversion}
\setboolean{publicversion}{true}

\ifthenelse{\boolean{publicversion}}{
   \newcommand{\grumbler}[2]{}
   
}{
   \newcommand{\grumbler}[2]{\MP{{\bf #1}: #2}}
    
}

\newmdenv[topline=false, bottomline=false, rightline=false,
          linewidth=2pt, linecolor=sumrule,
          backgroundcolor=sumbg,
          innerleftmargin=5pt, innerrightmargin=5pt,
          innertopmargin=3pt, innerbottommargin=3pt,
          skipabove=3pt, skipbelow=3pt]{keyideabox}
\newcommand{\keyidea}[2]{%
  \begin{keyideabox}\noindent\textbf{Key idea \##1:} #2\end{keyideabox}}
\newmdenv[topline=false, bottomline=false, rightline=false,
          linewidth=2pt, linecolor=idearule,
          backgroundcolor=ideabg,
          innerleftmargin=5pt, innerrightmargin=5pt,
          innertopmargin=3pt, innerbottommargin=3pt,
          skipabove=3pt, skipbelow=3pt]{evaltakeaway}
\makeatletter

\newcounter{takeaway}

\newcommand{\takeaway}[1]{%
  \refstepcounter{takeaway}%
  \@ifnextchar\bgroup{\mx@takeawayB{#1}}{\mx@takeawayA{#1}}%
}

\newcommand{\mx@takeawayA}[1]{%
  \begin{evaltakeaway}
    \noindent\textbf{Takeaway \#\thetakeaway:} #1
  \end{evaltakeaway}}

\newcommand{\mx@takeawayB}[2]{%
  \begin{evaltakeaway}
    \noindent\textbf{Takeaway \#\thetakeaway:} #2
  \end{evaltakeaway}}

\makeatother

\newmdenv[topline=false, bottomline=false, rightline=false,
          linewidth=2pt, linecolor=sumrule,
          backgroundcolor=sumbg,
          innerleftmargin=5pt, innerrightmargin=5pt,
          innertopmargin=3pt, innerbottommargin=3pt,
          skipabove=3pt, skipbelow=3pt]{summarybox}

\newboolean{appendix}
\setboolean{appendix}{true}

\usepackage{amsmath}
\usepackage[table, dvipsnames]{xcolor}
\usepackage{pifont}%
\usepackage{multirow}
\usepackage{algorithmic}
\usepackage[ruled,vlined,linesnumbered]{algorithm2e}

\usepackage{hyperref}
\usepackage{tcolorbox}
\tcbuselibrary{skins} %
\usepackage{multicol}

\usepackage{multibib}
\newcites{leg}{Legislation References}
\usepackage{cuted}

\usepackage{makecell}
\usepackage{circledsteps}

\newcommand{\projectname}{\textsc{Ember}\xspace}

\definecolor{darkpastelgreen}{rgb}{0.01, 0.75, 0.24}
\definecolor{darkpastelred}{rgb}{0.76, 0.23, 0.13}

\newcommand{\myparagraph}[1]{ \noindent{\bf {#1}.}}

\newcommand{\out}[1] {}

\newcounter{codeLineCntr}

\reversemarginpar
\newif\ifnotes
\notestrue

\newcommand{\punt}[1]{}

\renewcommand{\eqref}[1]{Equation~(\ref{eq:#1})}

\newcommand{\proc}[1]{\ifmmode\mbox{\textsc{#1}}\else\textsc{#1}\fi}

\newcommand{\func}[1]{\ifmmode\mathrm{#1}\else\textrm{#1}fi} %

\newcounter{remark}[section]

\usepackage[inline]{enumitem}
\setlist{leftmargin=16pt,noitemsep,topsep=0pt,parsep=0pt,partopsep=0pt}

\usepackage{setspace}
\usepackage[margin=5pt,font={stretch=0.9}]{caption}

\begin{document}

\date{}

\title{\projectname{}: Making Unified Virtual Memory Practical for Discrete-Memory GPUs}

\author{Senad-Leandro Lemes Galera}
\orcid{0009-0000-9496-1651}
\authornotemark[1]
\affiliation{%
    \institution{Technical University of Munich}
    \city{Munich}
    \country{Germany}
}
\email{senad.lemes@tum.de}

\author{Michail Georgoulakis Misegiannis}
\orcid{0009-0006-1106-9283}
\affiliation{%
    \institution{Technical University of Munich}
    \city{Munich}
    \country{Germany}
}
\email{geom@in.tum.de}

\authornote{Both authors contributed equally to this research.}

\author{Ferdinand Gruber}
\orcid{0009-0001-5703-3087}
\affiliation{%
    \institution{Technical University of Munich}
    \city{Munich}
    \country{Germany}
}
\email{gruberfe@in.tum.de}

\author{Jana Giceva}
\orcid{0000-0002-1926-3551}
\affiliation{%
    \institution{Technical University of Munich}
    \city{Munich}
    \country{Germany}
}
\email{jana.giceva@in.tum.de}

\author{Dimitrios Stavrakakis}
\orcid{0000-0002-3667-3763}
\affiliation{%
    \institution{Technical University of Munich}
    \city{Munich}
    \country{Germany}
}
\email{dimitrios.stavrakakis@tum.de}

\begin{abstract}
GPU working sets have outgrown GPU memory, so applications must \emph{oversubscribe} it, keeping only hot data on the GPU. NVIDIA's Unified Virtual Memory (UVM) supports \textit{oversubscription}, but its performance degrades once data exceeds the GPU's memory. This stems from its fault-driven data placement, wasting interconnect bandwidth, and underutilizing the GPU. 
This raises the question: {\em can UVM oversubscription become practical on stock GPUs, exploiting GPU memory speed and minimizing wasted interconnect bandwidth, without placement hints, profiling runs, or application changes?}

To this end, we present \sys, a UVM management framework for oversubscribed workloads. \sys introduces a \emph{hybrid address space}; allocations start as mapped into host memory, omitting data migration. \sys's \emph{counter-driven region promotion} uses GPU's hardware counters to fetch hot regions to the GPU, its \emph{decay-based eager reclamation} proactively evicts cold regions in batches, while via \emph{online region re-placement}, \sys tracks the hot set at runtime.

We implement \sys within NVIDIA's open-source UVM driver for unmodified applications and as a userspace variant on stock CUDA. At $1.5\times$ oversubscription, it is $6.8$--$10.6\times$ faster than UVM on six applications across three PCIe GPUs, and generates LLM tokens $4.7$--$16\times$ faster at sizes where manual placement cannot run at all. Without a profiling run, it beats a profile-guided alternative by $2.1\times$ on kernel time.

\end{abstract}

\maketitle
\pagestyle{plain}

\section{Introduction}
\label{sec:introduction}

GPUs have become the primary platform for data-intensive workloads~\cite{meta_grand_teton_2023, aws_accelerated_computing_2026, gcp_gpu_instances_2026, DBLP:conf/sigmod/InterlandiBHCSL26, top500_highlights_nov2025}. Their working sets are increasingly outgrowing GPU memory at a fast pace~\cite{DBLP:conf/sosp/KwonLZ0ZY0ZS23,DBLP:journals/corr/abs-2508-05029, DBLP:conf/eurosys/Sabet0G20}.
For instance, over the last few years, the largest models grew by ${\sim}1000\times$ while GPU memory \textit{only} by ${\sim}5\times$~\cite{DBLP:conf/sc/RajbhandariRRSH21}.
Provisioning larger GPUs, or scaling out across more of them~\cite{DBLP:conf/ipps/LavoreMADBS24}, raises monetary cost~\cite{cloudzero_gpu_pricing_2026} without removing this ceiling, and in shared clusters each tenant receives only a fraction of one GPU's memory~\cite{DBLP:conf/hipc/RaviNZB21, DBLP:conf/eurosys/StratiMK24, DBLP:conf/usenix/ChoiKJAJKA22}.
Therefore, applications must \emph{oversubscribe} GPU memory, keeping a subset of their data on the GPU and serving the rest from host memory over the interconnect, since otherwise they cannot run at all once their data exceeds device memory~\cite{nvidia_cuda_unified_memory_2026}.

\begin{figure}[t]
  \centering
  \includegraphics[width=\columnwidth]{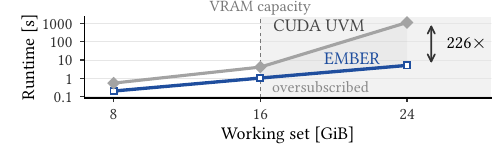}
  \vspace{-0.5cm}
  \caption{UVM performance collapse past the VRAM line (H200 capped to $16$\,GiB, GPU sort over scattered data).}
  \label{fig:collapse}
\end{figure}

To simplify GPU memory management, NVIDIA introduced Unified Virtual Memory (UVM)~\cite{nvidia_pascal_whitepaper_2016}, which places CPU and GPU memory in a single virtual address space and implicitly transfers data between them.
Since UVM pages data into the GPU on demand, an allocation may also exceed the device memory, allowing applications to oversubscribe the GPU.
Unlike manual data placement, which couples an application to a specific memory layout and requires rewriting per application and GPU generation~\cite{DBLP:journals/corr/abs-2508-05029, DBLP:conf/micro/RhuGCZK16, DBLP:journals/tpds/Nadal-SerranoL16}, UVM lets host and GPU share pointer-based structures~\cite{nvidia_um_performance, nvidia_cuda_runtime_memory_2026}.
By default, a GPU access to a non-resident page raises a page fault and the driver must \emph{migrate} the page into GPU memory before the access can complete, with later accesses served at device-memory bandwidth.
UVM also allows developers to \emph{prefetch} a range into the device, or \emph{map} it so the GPU reads host memory directly at interconnect speed.

Despite the benefits, UVM sees little adoption for working sets exceeding GPU memory, as performance collapses once the GPU capacity line is crossed. %
\autoref{fig:collapse} shows a GPU sort whose runtime grows from $4.1$\,s to $1150$\,s at $1.5\times$ oversubscription.
Thus, production systems either manually orchestrate the placement policies~\cite{DBLP:journals/corr/abs-2508-05029} or disable UVM~\cite{pytorch_uvm_rfc}, with vendors recommending steering placement by hand~\cite{nvidia_oversubscription_blog, nvidia_um_performance}.

Making UVM practical beyond device memory has four obstacles (\S\ref{sec:motivation}).
First, UVM suffers from \textit{reclamation-induced execution stalls} (\textit{C1}). 
Once the device memory is full, a page fault must also reclaim a device-resident page before the driver migrates the new one, and the reclaimed pages may become the next faults. 
This cycle runs on the critical path, stalling computation.
Architectural proposals shorten this path~\cite{DBLP:conf/isca/GangulyZYM19, DBLP:conf/hpca/ZhengNZSK16, DBLP:conf/micro/Ausavarungnirun17, DBLP:journals/taco/ZiabariSMSAUKJK16, DBLP:journals/corr/abs-2411-05309, DBLP:conf/asplos/0015ARZMGY19, DBLP:journals/tcad/YuCHQW20}, but they rely largely on simulations or on features absent from commodity GPUs.

Second, to reduce page faults, developers can map a memory range so that the GPU accesses host memory directly.
However, memory requests are then served over the interconnect, leading \textit{interconnect-bound mapped access} (\textit{C2}).
Thus, mapping alone cannot resolve UVM's performance issues.

Third, to leverage GPU memory performance, UVM allows the programmer to selectively prefetch data.
However, beyond the VRAM capacity, every transfer must first evict pages to free up space on the device. 
Prefetching moves up to ${\sim}1.5\times$ the amount of data that the default UVM does, trading faults for \textit{prefetch-amplified reclamation traffic} (\textit{C3}).
Hence, prefetching only defers UVM's performance collapse.

Fourth, while mapping and prefetching can be combined through placement hints, there is no principled rule specifying which to apply where, and a wrong hint can degrade performance~\cite{DBLP:conf/sc/ChienPM19, DBLP:journals/tjs/KnapC19,DBLP:conf/sc/Allen021}.
A hint also applies to a whole managed allocation and must be given before launch. 
However, access patterns shift during runtime~\cite{DBLP:conf/hpca/WangLJYT24, DBLP:conf/hpca/WangLZE0JT25}, leading to \textit{stale, allocation-granular placements} (\textit{C4}).
Prior work either fixes placement before the run and per allocation~\cite{DBLP:conf/sc/JonesAG25, DBLP:conf/micro/RhuGCZK16, DBLP:conf/sc/RajbhandariRRSH21}, or decides at runtime but still migrates by default, mapping only as a fallback, so reclamation stays on the critical path~\cite{DBLP:conf/hpca/ShinBPK26}.

\begin{mdframed}[
    backgroundcolor=rqbg,
    linecolor=black!70,
    linewidth=1pt,
    innerleftmargin=5pt,
    innerrightmargin=5pt,
    innertopmargin=2pt,
    innerbottommargin=2pt,
    skipabove=0pt,
    skipbelow=0pt
]
\textbf{Research question: }
\textit{Can UVM oversubscription become practical on stock GPUs, exploiting GPU memory speed, minimizing wasted interconnect bandwidth, without placement hints, profiling runs, or application changes?}
\end{mdframed}

To answer this question, we present \sys, a GPU memory management framework designed for oversubscribed workloads. %
Our key insight is that mapping should be the default, not a fallback.
Concretely, \sys decides which regions leave device memory and when, reclaiming cold regions in batches before memory fills, avoiding computation stalls (\textit{C1}).
Underneath, it builds a \emph{mapped-by-default hybrid address space}, in which every allocation starts as mapped and is reachable from the GPU without migration (\textit{C2}).
Only hot regions are promoted to GPU memory, either as advised by the developer in userspace or as reported by the GPU's hardware access counters in the kernel (\textit{C3}).
The split is also re-evaluated at runtime, demoting regions whose demand has decayed and promoting those that become hot (\textit{C4}).

We implement two variants of \sys: \emph{\sys-static}, a userspace variant on stock CUDA that fixes the data split at allocation time, and \emph{\sys-adaptive}, a patch to NVIDIA's open-source UVM kernel module that decides the split online and transparently supports unmodified applications.
We evaluate \sys on four GPUs, covering three common deployment settings, across three workload classes.
At $1.5\times$ oversubscription, it reaches a geometric-mean speedup of $6.8$--$10.6\times$ over naive UVM across six applications, sustains LLM generation at $4.7$--$16\times$ the throughput of UVM at model sizes exceeding the GPU memory capacity, and outperforms profile-guided placement by $2.1\times$ without a profiling run.

\myparagraph{Contributions} This paper makes four contributions:
\begin{itemize}[noitemsep,leftmargin=*,nosep]
\item \textbf{UVM performance analysis.} We show the page reclamation effect behind UVM performance issues, and that mapping or prefetching alone is not enough (\S\ref{sec:motivation}).
\item \textbf{Hybrid data placement.} We design a \textit{mapped-by-default hybrid address space} that assigns residency per region instead of per allocation, turning eviction from a permanent cost of oversubscription into a policy decision (\S\ref{sec:overview}, \S\ref{sec:design}).
\item \textbf{Counter-driven runtime data placement.} We derive the data split from the GPU's hardware access counters, and revise it while kernels run, demoting and promoting regions with no application changes (\S\ref{sec:design}, \S\ref{sec:impl}).
\item \textbf{Implementation \& evaluation.} We implement \sys's variants, and evaluate them using four discrete GPUs (\S\ref{sec:eval}).

\end{itemize}

\begin{figure}[t]
  \centering
  \includegraphics[width=\columnwidth]{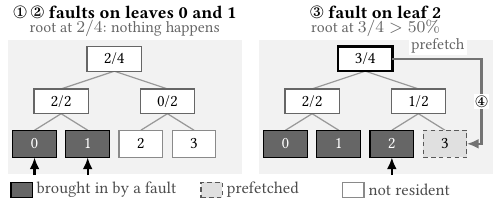}
  \vspace{-0.55cm}
  \caption{Default TBN prefetching mechanism~\cite{DBLP:conf/isca/GangulyZYM19}.}
  \label{fig:tbn}
\end{figure}

\section{Background}
\label{sec:background}
\subsection{Unified Virtual Memory (UVM)}
\label{sec:bg:uvm}
NVIDIA implements UVM in the \texttt{nvidia-uvm} kernel module, which manages the unified address space.
The driver divides each allocation into fixed-size $2$\,MiB \emph{VA blocks} and tracks, per block, which of its pages (e.g., $64$\,KiB pages on aarch64) are on the GPU. 
This is the unit at which placement is decided, and we call it a \emph{region}.
Physical GPU memory is allocated and reclaimed in \emph{chunks} of up to $2$\,MiB.
When the GPU accesses a page that is not resident, it faults into the driver and stalls for thousands of cycles until the page is migrated~\cite{DBLP:conf/ispass/VeselyBOLB16}.

\subsection{Data Placement}
\label{sec:bg:tbn}
\label{sec:bg:control}

There are three ways to place a page in UVM memory model.

\myparagraph{Migrate}
Migration through page faults is UVM's default strategy.
To amortize fault costs, CUDA prefetches speculatively with a \emph{tree-based neighborhood} (TBN) policy~\cite{DBLP:conf/isca/GangulyZYM19}.
Each $2$\,MiB VA block forms a tree whose leaves are its pages, and each internal node tracks how much of its subtree is GPU resident (\autoref{fig:tbn}).
A page fault makes its leaf resident, which raises a resident count at every node above it.
When more than half of a node's subtree is resident, the driver migrates the rest of it.
Thus, faults progressively migrate larger ranges, up to $2$\,MiB, betting on \emph{spatial locality}~\cite{DBLP:journals/tpds/ZhangWJWCCLL25}.

Under oversubscription, migrating a page requires freeing GPU memory.
UVM reclaims space with an age-based, approximately LRU policy that evicts chunks back to host memory over the same interconnect used for migration~\cite {DBLP:conf/isca/GangulyZYM19}.

\myparagraph{Prefetch}
A developer can move a range into the GPU ahead of use with \texttt{cudaMemPrefetchAsync}~\cite{nvidia_cuda_runtime_memory_2026, DBLP:conf/sc/ChienPM19}, avoiding many faults with one transfer.
Prefetching is still migration, however, so past capacity it must first free GPU memory.

\myparagraph{Map}
A developer can place a range in host memory, directly accessed by the GPU.
\texttt{cudaMemAdvise} attaches a placement policy to a range; \texttt{SetPreferredLocation} hints where it should live; and \texttt{SetAccessedBy} adds it to the GPU page table.
By combining these calls, the range can be preserved in the host (i.e., never migrated) and be available to the GPU without a fault. %
Such \emph{mapped} access only fetches the cache lines a kernel requests via the interconnect on every access.

Outside UVM, \emph{pinned explicit movement} stages data into pinned host buffers and copies it using \texttt{cudaMemcpy}, achieving high bandwidth but ruling out oversubscription, since the working set must fit in GPU memory.

\subsection{GPU Hardware Access Counters}
\label{sec:bg:counters}
NVIDIA GPUs expose roughly $256$ \emph{hardware access counters}, which count the GPU's accesses to host-resident memory at a configurable granularity~\cite{nvidia_cuda_runtime_memory_2026, DBLP:conf/isca/BSB26}.
Once a counter crosses its threshold, the GPU writes an entry into a buffer and keeps running, so unlike a fault it does not stall.
The driver reads these entries off the access path to see which range was accessed and how often, and we call each entry a \emph{notification}.
The counters are too few to cover a large mapped dataset, and the driver does not migrate based on them by default~\cite{DBLP:conf/isca/BSB26}.
The driver also discards notifications when the operating system's page size is $4$\,KiB, the default on x86 Linux, because the migration path they would drive is unsupported at that size~\cite{nvidia_open_gpu_kernel_modules}.
The counters therefore go unused on most systems.

\begin{figure}[t]
  \centering
  \includegraphics[width=\columnwidth]{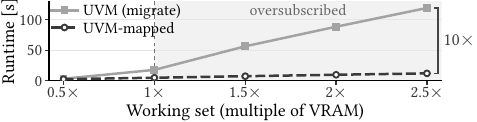}
  \vspace{-0.55cm}
  \caption{Runtime of UVM and UVM-mapped (A40, random access, 8\,B elements).}
  \label{fig:decomp}
\end{figure}

\begin{figure}[t]
  \centering
  \includegraphics[width=\columnwidth]{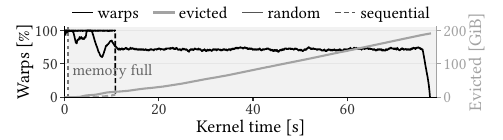}
  \vspace{-0.55cm}
  \caption{Warps in flight and cumulative eviction (right axis) under oversubscription (RTX~5080, 10\,ms samples).}
  \label{fig:warp}
\end{figure}

\subsection{Coherent Interconnects}
\label{sec:bg:coherent}
Recent systems connect the CPU and GPU over a cache-coherent interconnect (e.g., NVLink-C2C on Grace Hopper~\cite{nvidia_grace_hopper_2026, DBLP:conf/icpp/SchiefferW0FP24}), where they share one page table.
The GPU accesses host memory directly, and the gap between device-memory and interconnect bandwidth is lower compared to PCIe (\autoref{tab:hw}).
However, GPU memory remains a separate, finite pool, and a resident set must still be chosen. %
Finally, physically unified memory promises to remove the distinction between host and device memory altogether~\cite{apple_m1_2020}.

\section{Technical Challenges and Key Ideas}
\label{sec:motivation}

UVM enables GPU memory oversubscription without data-movement code.
However, under oversubscription, its three placement mechanisms and their combinations fail.

\myparagraph{C1: Reclamation-induced execution stalls}
\label{subsec:mot:tbn}
Hiding data movement is a core benefit of UVM, but its TBN prefetcher and age-based eviction thrash once the working set exceeds VRAM (\S\ref{sec:bg:tbn}).
Past the VRAM line, UVM runs up to $10\times$ slower than mapping the same data from the host (\emph{UVM-mapped}), which stays flat even though it is interconnect-bound (\autoref{fig:decomp}).
\emph{UVM-mapped} never incurs a page fault and never migrates data, so nothing is ever reclaimed, which helps locate UVM's cost.
A UVM kernel with random accesses runs $7\times$ longer than one with sequential accesses, even though it incurs $5\times$ \emph{fewer} page faults, because it evicts $26\times$ more data.
TBN explains the difference, since it prefetches a neighborhood for each fault, which a sequential kernel consumes and never needs again, whereas a random kernel uses little of it and evicts pages before it accesses them again.
\autoref{fig:warp} shows what that costs.
Once memory is full, eviction never stops, and because each eviction interleaves with execution and stalls the warps whose pages it takes, warp occupancy falls and never recovers.

\begin{figure}[t]
  \centering
  \includegraphics[width=\columnwidth]{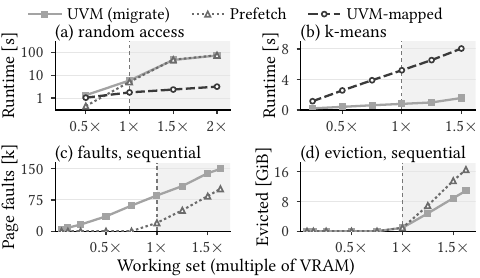}
  \vspace{-0.35cm}
  \caption{Allocation strategies across the VRAM line (A40): (a)~random access; (b)~arithmetic-intensive k-means; (c)~fault rate and (d)~evicted volume under prefetching.}
  \label{fig:pure}
\end{figure}

\begin{figure*}[t]
  \centering
  \includegraphics[width=\textwidth]{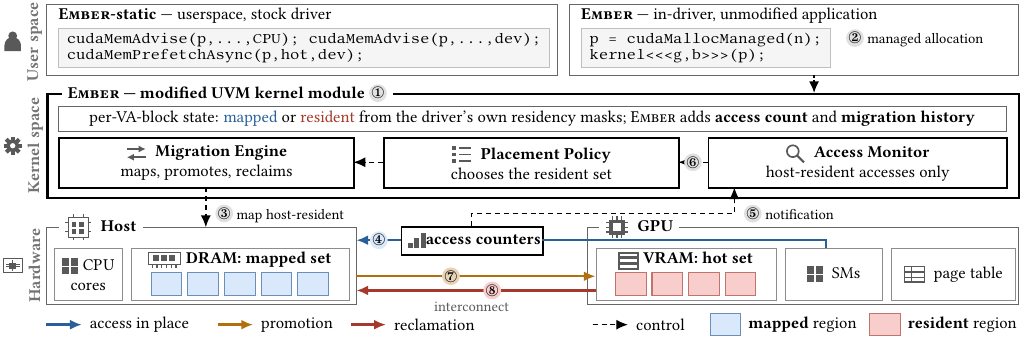}
  \vspace{-0.5cm}
  \caption{\sys architecture and allocation workflow.
  Top left: \sys-static, the userspace stand-in on the stock driver.}
  \label{fig:overview}
  \vspace{-0.2cm}
\end{figure*}

\keyidea{1}{To avoid reclamation traffic and page faults, \sys maps allocations in host memory by default (\S\ref{sec:design:engine}).}

\myparagraph{C2: Interconnect-bound mapped access}
\label{subsec:mot:zerocopy}
While \emph{UVM-mapped} does not reclaim any data and can even improve performance in some scenarios (\autoref{fig:pure}a), it leaves device-memory bandwidth unused.
Every access is bounded by the interconnect, which has $15$--$150\times$ lower bandwidth than device memory over PCIe and still ${\sim}9\times$ lower over NVLink-C2C (\autoref{tab:hw}).
Compute-intensive kernels are affected most, since every reuse of their data crosses the interconnect.
For instance, a GPU-accelerated k-means application runs up to $6.5\times$ slower when served purely from the host (\autoref{fig:pure}b), and LLM workloads are bounded in a similar fashion (\S\ref{sec:eval-llm}).

\keyidea{2}{To recover device-memory bandwidth, \sys splits each allocation, serving selected regions from device memory and leaving the rest mapped on the host (\emph{mapped-by-default hybrid address space}, \S\ref{sec:design:engine}).}

\myparagraph{C3: Prefetch-amplified reclamation traffic}
\label{subsec:mot:prefetch}
Prefetching large data ranges ahead of use (\emph{Prefetch}) fills the GPU, leaving no room for subsequent data fetches, once the VRAM capacity is reached.
On random accesses, Prefetch collapses similarly to stock UVM (\autoref{fig:pure}a).
It falls back to on-demand faulting at UVM-like fault rates (\autoref{fig:pure}c), but evicts up to ${\sim}1.5\times$ the data of stock UVM (\autoref{fig:pure}d).
Thus, prefetching replaces faults with bulk transfers but increases the eviction volume and, as a consequence, the reclamation traffic~\cite{DBLP:conf/ipps/Allen021}.

\keyidea{3}{To avoid reintroducing eviction thrashing, \sys promotes only the regions the GPU's access counters show as hot, and stops promoting before device memory fills (\emph{access-counter-driven region promotion}, \S\ref{sec:design:monitor}).}

\begin{figure}[t]
  \centering
  \includegraphics[width=\columnwidth]{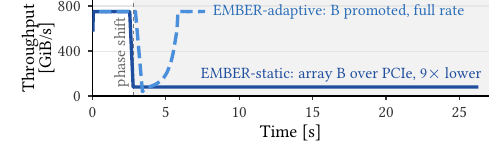}
  \vspace{-0.55cm}
  \caption{Throughput across a phase shift, split fixed or re-derived online (RTX~5080, array~A then array~B; \S\ref{sec:eval-ablation}).}
  \label{fig:phase}
\end{figure}

\myparagraph{C4: Stale, allocation-granular placement}
\label{subsec:mot:static}
Combining mapping and selective prefetching can help, but it is not enough, because the split between them must be both fine-grained and adjustable at runtime.
Specifically, an allocation mixes hot and cold data~\cite{DBLP:journals/tpds/ZhangWJWCCLL25}.
A decision at the allocation granularity can lead either to promoting cold data to VRAM, occupying useful space, or to leaving hot data in host memory.
Which data is hot also changes as an application moves between phases.
With a split fixed before launch, a kernel that reads array~A and then array~B on a footprint larger than GPU memory loses ${\sim}9\times$ of its throughput at the switch, from $750$ to $80$\,GiB/s, because device memory still holds~A.
A split re-derived online recovers instead (\autoref{fig:phase}).

\keyidea{4}{To keep the split current as the hot set moves, \sys reclaims regions whose demand has decayed, in batches before GPU memory fills, and promotes those that turn hot again (\emph{decay-based eager reclamation} and \emph{online region replacement}, \S\ref{sec:design:policy}).}

\section{Overview}
\label{sec:overview}

\sys is a GPU memory management framework tailored for oversubscribed workloads.
It aims to \emph{(i)}~benefit from GPU memory performance while the working set fits, \emph{(ii)}~avoid the collapse on oversubscription, and \emph{(iii)}~require no changes to hardware, GPU kernels, or applications.

\subsection{System Architecture}
\label{sec:overview:arch}
 
We build two variants of \sys:
\emph{(i)}~\emph{\sys-static} establishes the hybrid address space from user space with placement hints and fixes the data split between host and device before kernel launch, for hot sets known up front and for machines without bare-metal access (e.g., cloud instances).
\emph{(ii)}~\emph{\sys-adaptive} executes inside the UVM driver and adjusts this split at runtime from the GPU's hardware access counters, for hot sets that are unknown or change during the run (\S\ref{sec:eval-overhead}).

\autoref{fig:overview} shows \sys's architecture for both variants.
Its core consists of three components: a \emph{migration engine}, an \emph{access monitor}, and a \emph{placement policy engine}.
The \emph{migration engine} owns the \emph{mapped-by-default hybrid address space}, in which mapped and device-resident regions coexist within one allocation, and carries out every change of placement (\autoref{sec:design:engine}).
The \emph{access monitor} turns access-counter notifications into per-region demand without consuming GPU cycles, thereby enabling \emph{counter-driven region promotion} (\autoref{sec:design:monitor}).
\sys needs little state of its own, since UVM's per-VA-block state already records where a region resides and \sys tracks only how often it has moved and how much demand the counters last reported (\autoref{sec:design:state}).
The \emph{placement policy engine} consumes that demand and decides what moves, through \emph{decay-based eager reclamation} and \emph{online region replacement} (\autoref{sec:design:policy}).

\subsection{Life of an Allocation}
\label{sec:overview:life}

\autoref{fig:overview} presents a managed allocation in \sys.
Each region is either \emph{mapped} or \emph{device-resident}.

\begin{lstlisting}[float=t, belowskip=-3pt, style=mystyle, basicstyle=\ttfamily\footnotesize, caption={\sys-static needs three calls per allocation on the stock driver; \sys-adaptive lifts this requirement.}, label={lst:integration}]
// EMBER-static: userspace, stock driver
cudaMallocManaged(&ptr, n);
|\Hilightt|cudaMemAdvise(ptr, n, SetPreferredLocation, CpuDeviceId);
|\Hilightt|cudaMemAdvise(ptr, n, SetAccessedBy, dev);
|\Hilightt|cudaMemPrefetchAsync(ptr, hot_size, dev); // split is fixed
kernel<<<g,b>>>(ptr);

// EMBER-adaptive: in-driver, unchanged application
cudaMallocManaged(&ptr,n);
kernel<<<g,b>>>(ptr);
\end{lstlisting}

\myparagraph{Deployment phase}
To deploy \sys-static, a developer adds three placement calls after each \texttt{cudaMallocManaged} (\autoref{lst:integration}), which establish the hybrid address space and fix the split at launch.
\sys-adaptive needs only the kernel module to be loaded (\stepc{1}) and leaves the application intact.

\myparagraph{Allocation phase}
Every managed allocation reaches \sys without an API call (\stepc{2}), and the migration engine maps it, with the host memory as its preferred location (\stepc{3}).
As in stock UVM, pages are populated lazily. 
A page first accessed by the host is created in host memory and mapped; one first touched by the GPU has no copy to move yet, so \sys reads the fault as demand and creates the page in device memory at no transfer cost (\S\ref{sec:design:monitor}).
A kernel can therefore start without waiting for any data to move, even when the allocation is many times larger than the device memory (\stepc{4}).

\myparagraph{Promotion phase}
Under \sys-adaptive, every region starts mapped, so the GPU reads it over the interconnect while the hardware access counters record its accesses.
The access monitor collects notifications from the access counters and connects each to their VA block. 
The counting happens in hardware, without costing GPU time.
When an access count crosses a threshold (\stepc{5}), the placement policy engine marks it to be promoted (\stepc{6}), and the migration engine locks the VA block, migrates its host-resident pages and updates the page table (\stepc{7}).
After this point, the region is \emph{device-resident}, and since the counters capture only host memory, its recorded demand only decays from here on (\S\ref{sec:design:policy}).

\myparagraph{Reclamation phase}
Before the GPU memory fills, the policy ranks resident regions by their migration history and by their recorded demand before promotion, and writes the coldest back in batches (\stepc{8}) to provide space for upcoming promotions without evicting in the critical path.
A reclaimed region returns to host memory, where the access counters see it again, so a region evicted by mistake is promoted back as soon as the GPU keeps accessing it.

\subsection{System Model}
\label{sec:overview:model}

\myparagraph{Deployment model}
\sys targets discrete GPUs, whose device memory is oversubscribed.
\sys-static runs on the stock driver without requiring elevated privileges or bare-metal access on the host.
A developer deploys it using placement hints (\autoref{lst:integration}).
It preserves the UVM memory semantics but fixes the split before launch, preventing it from following the access patterns of long-running applications.
\sys-adaptive lifts this restriction by making placement decisions at runtime via a kernel module that extends NVIDIA's UVM module.
It requires hardware access counters, standard in most GPUs, and a machine on which an administrator can load a module (\S\ref{sec:eval-limits}).
\sys handles all managed allocations, so unmodified binaries can seamlessly adopt it, with the policy parameters (\autoref{tab:params}) set at module load time.

\myparagraph{Programming model}
\sys requires \textit{no change} to the UVM programming model.
An application allocates managed memory and dereferences the returned pointer in both host and device code, as before.
Managed memory keeps its semantics, the coherence guarantees of stock UVM, and allocations may exceed device memory.
\sys only \textit{transparently} determines where each page resides during execution. 
Currently, allocations outside the \sys-managed path, such as GPU-only buffers and buffers created within third-party libraries, retain their existing semantics and remain outside \sys's control (\S\ref{sec:eval-limits}).
An application that issues its own placement hints on a managed range overrides \sys for that range, so application-specific optimizations are preserved.

\section{Design}
\label{sec:design}

\sys realizes its key ideas (\autoref{sec:motivation}) via three core components:
the \textit{migration engine} provides the hybrid address space and carries out data movement (\autoref{sec:design:engine}), the \textit{access monitor} produces the placement hints (\autoref{sec:design:monitor}), and the \textit{placement policy engine} consumes them to make placement decisions (\autoref{sec:design:policy}).

\subsection{Migration Engine}
\label{sec:design:engine}
\label{sec:design:migration}
The migration engine owns the address space and manages physical data placement, realizing the \emph{mapped-by-default hybrid address space} and \sys's block promotion mechanism.

\myparagraph{Hybrid address space}
\label{sec:design:address-space}
\sys divides each managed allocation into more granular units (regions), which, in a steady state, are either \emph{mapped} or \emph{device-resident} (\autoref{sec:overview}).
Mapped regions indicate the performance \emph{floor} as they are immediately available once mapped, never faulted in and evicted again, and bounded, in the worst case, by interconnect bandwidth. 
On the other hand, device-resident regions provide a performance \emph{ceiling}, since a resident region is served at GPU memory bandwidth for as much data as the GPU can hold.

\begin{figure}[t]
  \centering
  \includegraphics[width=\columnwidth]{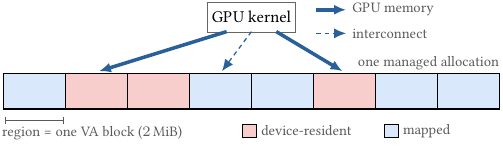}
  \vspace{-0.4cm}
  \caption{\sys's hybrid address space layout.}
  \label{fig:hybrid}
\end{figure}

\sys allows regions in both states to coexist within a single allocation (\autoref{fig:hybrid}).
A region is mapped as soon as its pages exist and remains mapped until it earns promotion, inverting the stock UVM arrangement, where fetching data on the GPU is the default and mapping is the exception.
This inversion makes any allocation immediately usable, starts kernel execution without data movement, and defers reclamation until the policy engine initiates it.

\myparagraph{Region state}
\label{sec:design:state}
\sys defines a region to be exactly one VA block, and extends the driver's per-VA-block state with a migration history, and attaches a decayed demand~\cite{DBLP:conf/sosp/LeeMME23} to each physical chunk backing a resident region (\autoref{sec:design:adaptive}).
As an alternative, \sys could group several VA blocks into a larger region to reduce the number of migration operations. 
However, since each block carries a lock, a single placement change would need to acquire multiple locks and force the migration path to reason about lock ordering and re-entrancy.
Instead, inheriting the driver's granularity preserves every invariant it maintains over a VA block, at the cost of a unit of state fixed by the driver, and a larger region would have to be re-established against each new driver release (\S\ref{sec:eval-limits}).

\sys recovers the batching benefit that a larger region would have given by promoting a configurable span of consecutive blocks at once (\autoref{sec:design:monitor}).
Reclamation works in physical chunks rather than blocks because chunks are the unit in which the driver hands back device memory.

\myparagraph{Region promotion/demotion}
When \sys initiates a region promotion, the migration engine locks its VA block, moves every host-resident page into GPU memory in one operation, and updates the GPU page table.
Thus, each promoted block costs \textit{only} one bandwidth-saturating transfer.

Demotion is the inverse operation.
When \sys performs a region demotion, the migration engine gathers a set of physical chunks and writes them back to host memory as a bounded batch. 
\sys preserves this operation off the critical path to avoid interleaving with computation, removing the observed stalls under the stock UVM reclamation policy (\autoref{fig:warp}).
The reclaimed region itself stays mapped and accessible over the interconnect, and the freed capacity returns to the driver's GPU memory allocator, where other clients can use it.
Coalescing a batch into a single device-to-host transfer and finer-grained triggers (e.g., idle-interconnect windows) are orthogonal and remain future work.

\myparagraph{Synchronization}
The migration engine holds at most one VA-block lock at a time.
A promotion locks only the block it moves, and a demotion runs as a separate pass. 
In this way, threads do not wait on a second lock, thereby avoiding deadlocks.
Within a migration, \sys calls the driver's own routine and holds that lock for its whole duration, including the page table update, as the routine requires.
The recorded residency of a block therefore never disagrees with where its pages are, and other threads see the new state only once the move has finished.
Access counter updates require no locks, since they serve only as hints. 
A lost or stale count delays a promotion by one interval without compromising correctness, with the block state staying valid.

\subsection{Access Monitor}
\label{sec:design:monitor}
The access monitor translates GPU memory traffic into data-placement signals, enabling \emph{counter-driven region promotion}.
Using the GPU's hardware access counters, it decides when a mapped region should be moved into GPU memory.

\myparagraph{Hardware access counters}
The UVM driver exposes hardware access counters that report how often a GPU reaches a host-resident range. While UVM does not use them and migrates on faults, \sys infers data placement hints based on them, attributing each one to the VA block whose pages were accessed.
This costs no GPU time and never blocks the kernel.
However, GPU counters have limitations.

\myparagraph{Limited counter view}
The access counters monitor only host memory~\cite{DBLP:conf/isca/BSB26}.
They cannot report the access frequency of GPU-resident regions or identify cold data suitable for eviction. 
\sys infers coldness in its policy engine instead, from the demand each region showed before promotion (\autoref{sec:design:adaptive}).

\myparagraph{Limited counter number}
A GPU has $\approx256$ counters~\cite{DBLP:conf/isca/BSB26}, that cannot cover a mapped set of hundreds of GiB at the VA-block granularity.
\sys compensates for this limitation by widening the scope of the action each counter notification triggers.
A notification is tied to one VA block, but a promotion covers a configurable span of consecutive blocks around it, as a sequence of independent per-block promotions (\autoref{sec:design:engine}).
Thus, in \sys, one observed block stands in for its neighborhood, and a few hundred counters can drive placement decisions across a mapped set that is much larger.

\begin{figure}[t]
    \centering
    \includegraphics[width=\columnwidth]{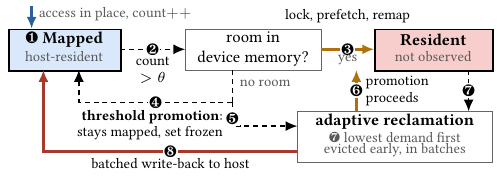}
    \vspace{-0.45cm}
    \caption{A region's lifecycle under \sys's policies.}
    \label{fig:lifecycle}
\end{figure}

\myparagraph{Promotion threshold}
\sys's decisions are based on a threshold that determines when a region earns promotion.
It encodes a break-even point, reached once a region has been frequently accessed, where serving further accesses from GPU memory outweighs the migration cost.
This threshold is crucial; if set too low, \sys may promote regions that are not reused, while high values might leave hot data on the host, repeatedly paying for access over the interconnect. Therefore, \sys exposes it as a tunable parameter. %

\myparagraph{Fault-driven promotion}
A region can also be promoted without a counter notification.
A mapped region never faults, so a fault occurs only on a page that has not yet been populated: the host has not accessed it, and there is no copy to map.
\sys handles it by creating a page in GPU memory at no transfer cost.
Thus, data a kernel produces starts resident while data the host produced starts mapped (\autoref{sec:overview}).

\subsection{Placement Policy Engine}
\label{sec:design:policy}
The access monitor provides placement hints, while the policy engine contains the rules for enforcing them.
Different policies can be applied without changes to the migration path or the application.
\sys implements two policies (\autoref{fig:lifecycle}), \emph{threshold promotion} and \emph{decay-based eager reclamation}.

\myparagraph{Threshold promotion}
\label{sec:design:freeze}
Under this policy, a mapped region is served from host memory while its counter accumulates (\stepn{1}); crossing a threshold $\theta$ makes it a promotion candidate (\stepn{2}), and \sys migrates it to GPU, if there exists available memory (\stepn{3}).
Note that eviction is \textit{disabled}.
Once the GPU memory fills, \sys freezes the resident set and leaves remaining regions in the host (\stepn{4}).
Thus, \sys benefits from GPU memory bandwidth for the initially accessed data, incurs no reclamation traffic, and resorts to interconnect-bound accesses for data beyond the GPU memory capacity. 

\myparagraph{Decay-based eager reclamation}
\label{sec:design:adaptive}
Despite the potential benefits of the static \emph{threshold promotion}, a frozen GPU-resident dataset can become stale as the hot dataset shifts.
For this reason, \sys also includes a \emph{decay-based eager reclamation} policy. 
Inspired by prior work~\cite{DBLP:conf/asplos/MarufWDWABPCKC23}, \sys applies eager reclamation and orders potential eviction candidates by decayed demand.
It frees space before demand arrives (\stepn{5}), removing eviction from the hot path.
Thus, it allows upcoming promotions to proceed (\stepn{6}), with victims inferred from decayed demand (\stepn{7}) and written back in batches (\stepn{8}).

Above a GPU occupancy threshold, \sys reclaims until it falls back to a lower one.
In this way, space for the next promotion becomes available before its notification arrives, and the computation can avoid the memory reclamation and eviction stalls on the hot path.
Below the upper threshold, it replaces GPU-resident regions that are colder than the hottest one by a configurable ratio.
The gap between the two thresholds prevents promotion and reclamation from clashing, as reclaiming down to the lower threshold leaves room for several promotions before the upper one is reached again.
A promotion that finds no free space fails and retries on the region's next notification, leaving reclamation as the only path to evict GPU-resident data (\autoref{sec:impl}).

\myparagraph{Demand decay}
Choosing eviction candidates requires knowing which GPU-resident regions have gone cold, which the hardware counters cannot report (\autoref{sec:design:monitor}).
\sys therefore records the demand \emph{before} it disappears, adding the access count of every serviced notification to the physical chunk backing a region.
Thus, a region arrives in GPU memory already carrying a measure of its past access frequency, which, then, halves at fixed time intervals.
This means that a moved region sinks to the coldest position on its own over time.

\myparagraph{Online region replacement}
Demotion returns a region to the mapped state.
A region reclaimed while it was still hot re-earns promotion within one notification interval, at the cost of an interconnect round trip.
A region undergoing migration is excluded from placement decisions until its migration completes.
The policy engine skips the regions the driver marks as in-flight and handles each promotion atomically; two service threads cannot start the same promotion twice.

\myparagraph{Migration-aware protection}
\sys's decay mechanism might eventually cool a region that is still genuinely hot.
To address this, a region that returns to the GPU shortly after being reclaimed has its \emph{migration score} increased, granting its underlying physical chunks a \textit{bounded} residency grace period proportional to that score.
The score decays more slowly than the demand, keeping a contested region protected across several reclamation passes.
Reclamation never evicts a protected region while an unprotected one exists. 
If all regions are protected, \sys selects the first to expire, ensuring that its protection can delay reclamation but never stalls an allocation.
Victims are ordered lexicographically, unprotected before protected, ranked by lowest migration score, oldest residency, and lowest decayed demand.

\myparagraph{Capacity accounting}
Capacity accounting is conservative, as promotion stops at the upper occupancy threshold that triggers reclamation.
The margin below full memory shows the accounted headroom.
It also keeps GPU memory from filling, so the stock fault-and-evict path never runs, neither under \sys-adaptive nor with \sys-static's prefix (\textbf{C1}).
That headroom is important as it covers what \sys cannot observe.
For instance, a GPU's free-memory figure can overstate what is usable, and allocations outside the managed path consume GPU memory invisibly to \sys (\autoref{sec:eval-limits}).

\section{Implementation}
\label{sec:impl}

\sys-static is a userspace component that runs on stock CUDA with an unmodified UVM driver.
\sys-adaptive, following prior work~\cite{DBLP:conf/iiswc/ZhuCVHCR22, DBLP:conf/systor/ChangKS21, DBLP:conf/isca/BSB26, DBLP:conf/hpca/ShinBPK26}, is a patch to the UVM module of NVIDIA's open-source kernel modules (\textit{595.71.05}), built on CUDA~\textit{12.4} and Linux~\textit{6.17}, adding 1194 and removing 31 lines across 11 files.
\autoref{tab:params} lists the module parameters and their default values, used in our \sys-adaptive prototype along with \texttt{cpu\_preferred\_blocks\_enable}.

\begin{table}[t]
\centering
\scriptsize
\caption{\sys module parameters; new ones in \textcolor{patchparam}{blue}.}
\label{tab:params}
\vspace{-0.1cm}
\setlength{\tabcolsep}{2pt}
\renewcommand{\arraystretch}{1.12}
\begin{tabular*}{\linewidth}{m{0.53\columnwidth}m{0.34\columnwidth}>{\centering\arraybackslash}m{0.08\columnwidth}}
\hline
    Parameter & Governs & Default \\
    \hline
    \multicolumn{3}{@{}l@{}}{\textbf{Promotion}}\\
    \textcolor{patchparam}{\texttt{cpu\_preferred\_blocks\_enable}}          & master switch              & 0 \\
    \textcolor{patchparam}{\texttt{cpu\_preferred\_block\_promotion\_enable}}& whole-block promotion      & 1 \\
    \textcolor{patchparam}{\texttt{cpu\_preferred\_prefetch\_chunk\_mb}}     & span promoted, MiB         & 64 \\
    \texttt{perf\_access\_counter\_threshold}                                & accesses before notifying  & 256 \\
    \multicolumn{3}{@{}l@{}}{\textbf{Reclamation} (prefix: \texttt{perf\_eager\_eviction\_})}\\
    \textcolor{patchparam}{\texttt{enable}}          & master switch              & 1 \\
    \textcolor{patchparam}{\texttt{high\_watermark}} & usage forcing eviction, \% & 90 \\
    \textcolor{patchparam}{\texttt{low\_watermark}}  & usage targeted, \%         & 85 \\
    \textcolor{patchparam}{\texttt{interval\_ms}}    & aging period, ms           & 25 \\
    \textcolor{patchparam}{\texttt{batch\_count}}    & chunks evicted per pass    & 16 \\
    \textcolor{patchparam}{\texttt{heat\_ratio}}     & hot:cold ratio to replace  & 2 \\
    \textcolor{patchparam}{\texttt{stale\_intervals}}& silent periods until cold  & 5 \\
    \multicolumn{3}{@{}l@{}}{\textbf{Migration-aware protection} (prefix: \texttt{perf\_migration\_})}\\
    \textcolor{patchparam}{\texttt{aware\_eviction\_enable}}     & master switch                & 1 \\
    \textcolor{patchparam}{\texttt{eviction\_base\_grace\_ms}}   & grace after arrival, ms      & 100 \\
    \textcolor{patchparam}{\texttt{eviction\_return\_window\_ms}}& window counting returns, ms & 500 \\
    \textcolor{patchparam}{\texttt{eviction\_score\_grace\_ms}}  & grace per score point, ms    & 250 \\
    \textcolor{patchparam}{\texttt{eviction\_max\_grace\_ms}}    & grace cap, ms                & 2000 \\
    \textcolor{patchparam}{\texttt{eviction\_decay\_ms}}         & score halving period, ms     & 1000 \\
    \hline
\end{tabular*}
\end{table}

\subsection{\sys-adaptive}
\label{sec:impl:adaptive}

\myparagraph{Memory allocation}
On a \texttt{cudaMallocManaged} call, \sys creates the VA range, sets its preferred location to the CPU, and adds every GPU to its accessed-by mask.
Pages are still populated lazily, as in stock UVM. 
The call itself moves no data, and \sys decides only where each page lands when it is first touched, without any input from the application.

A GPU fault can still occur on a page with no host copy (e.g., kernel's output).
\sys places that page in GPU memory with no recorded demand, making it a prime eviction candidate. 
After a potential reclamation, it remains mapped and is promoted if it proves to be hot, per \sys's policies.

\myparagraph{Monitoring}
\sys employs access counters for managed ranges and handles their notifications via the driver's existing path.
Each notification is associated with a VA block. 
The block's counter value is added to the decayed demand of the physical root chunk holding the block's pages, under a short lock on the chunk list of the driver's GPU memory allocator.
Thus, demand lives on physical chunks, while the VA block carries only its migration history and a flag marking an in-flight promotion.
\sys also lifts the $4$\,KiB-page restriction of GPU counters (\autoref{sec:bg:counters}) for ranges under its own policy.

\myparagraph{Promotion}
When a counter notification arrives, \sys locks the block and promotes it as a whole, swapping the notification's accessed-page mask with the block's host-resident mask and forcing the destination to the GPU.
\sys reuses the driver's migration routine, moving the block's host-resident pages to GPU memory and updating the GPU page table in a single operation.
\sys extends the promotion range to a span around the block, $64$\,MiB by default, centered on it except near an allocation boundary, where it shifts to accommodate the span size.
A span promotion skips the pages the driver protects and stops at the reclamation threshold, so it never displaces data.
A promotion halted for lack of space resumes at the region's next notification.
\sys records these events and the bytes moved, so the policy can be tuned without a profiler.

\begin{table}[t]
  \centering
  \footnotesize
  \caption{Evaluation platforms. The \emph{Gap} column indicates the device-memory-to-interconnect bandwidth ratio.}
  \label{tab:hw}
  \vspace{-0.1cm}
  \setlength{\tabcolsep}{1.3pt}
  \begin{tabular}{@{}lccccr@{}}
    \hline
    GPU & GPU memory & Interconnect & CPU & DRAM & Gap \\
    \hline
    RTX~5080 & 16\,GiB GDDR7  & PCIe~5.0 $\times16$   & Ryzen~7 9700X & 32\,GB & ${\sim}15\times$ \\
    A40      & 48\,GiB GDDR6  & PCIe~4.0 $\times16$   & Xeon Gold 6326 & 337\,GB & ${\sim}22\times$ \\
    H200     & 141\,GiB HBM3e & PCIe~5.0 $\times8$   & Xeon 6980P & 1\,TB & ${\sim}150\times$ \\
    GH200    & 96\,GiB HBM3   & NVLink-C2C & Grace & 480\,GB & ${\sim}9\times$ \\
    \hline
  \end{tabular}
\end{table}

\myparagraph{Reclamation}
\sys reclaims memory through a periodic background worker in the driver's GPU memory allocator.
Every $25$\,ms it halves each root chunk's demand and then selects victims.
Above $90\%$ occupancy it evicts until usage falls to $85\%$.
Below that point it evicts only when the hottest resident chunk carries at least twice the demand of the coldest.
A resident chunk whose demand has stayed at zero for five intervals is also treated as cold.
Each pass evicts at most 16 chunks and returns them to the driver's allocator.

\myparagraph{Migration-aware protection}
To prevent regions from bouncing between host and GPU memory, \sys records the migration events in the VA block's migration history.
A block that returns to the GPU within $500$\,ms of an eviction has its migration score increased by 1 for each $2$\,MiB returned. This score halves every second.
On arrival, the score sets a grace period on the chunks backing the block, $100$\,ms plus $250$\,ms per score point, capped at $2$\,s.
The eviction candidate search then proceeds, respecting the region state ordering (\autoref{sec:design:policy}).

\subsection{\sys-static}
\label{sec:impl:static}
\sys-static requests the initial placement from userspace and freezes the split.
At each allocation site, the developer issues two \texttt{cudaMemAdvise} calls, naming the host as the preferred location and adding the GPU to the accessed-by mask, followed by one \texttt{cudaMemPrefetchAsync} (\autoref{lst:integration}).
The prefetch migrates the leading prefix of the allocation, or a region equal to what \texttt{cudaMemGetInfo} reports free, to GPU memory, leaving the rest in host memory.

\sys-static is beneficial for applications that lay out their hot data contiguously and whose data split does not need to be revised at runtime.
Since \sys-static does not perform reclamation, it does not require counters. It makes only three calls per managed allocation (\autoref{lst:integration}) and preserves the kernel, launch code, and associated algorithms. We quantify the porting effort in Lines of Code (LoC) in \autoref{app:loc}.

\begin{figure*}[t]
  \centering
  \includegraphics{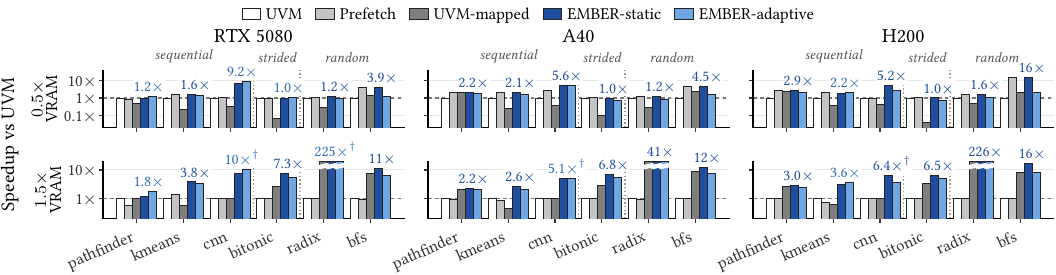}
  \vspace{-0.55cm}
  \caption{Applications at $0.5\times$ (top) and ${\sim}1.5\times$ (bottom) device memory, speedup over stock UVM (log scale). Labels give the faster \sys variant, colored by variant. Dagger: stock-UVM baseline hit the $1800$\,s cap.}
  \label{fig:bar}
  \vspace{-0.5cm}
\end{figure*}

\begin{table}[b]
  \footnotesize
  \centering
  \caption{Evaluation variants.}
  \vspace{-0.1cm}
  \label{tab:variants}
  \setlength{\tabcolsep}{3pt}
  \begin{tabular}{@{}l>{\raggedright\arraybackslash}m{5cm}c@{}}
    \hline
    Variant & Data Placement & APIs \\
    \hline
    UVM           & default with TBN~\cite{DBLP:conf/isca/GangulyZYM19} & M \\
    Prefetch      & bulk migration at boot & M+P \\
    UVM-mapped    & mapped from host DRAM & M+A \\
    HELM~\cite{DBLP:conf/sc/JonesAG25} & per-allocation choice after profiling & - \\
    \hline
    \sys-static   & hot (GPU)/cold (DRAM) fixed split & M+A+P \\
    \sys-adaptive & decided online based on access counters & M\\
    \hline
    \multicolumn{3}{@{}l}{\scriptsize \textbf{M}: \texttt{cudaMallocManaged}; \textbf{P}: \texttt{cudaMemPrefetchAsync}; \textbf{A}: \texttt{cudaMemAdvise}} \\
  \end{tabular}
\end{table}

\begin{figure}[t]
  \centering
  \includegraphics[width=\columnwidth]{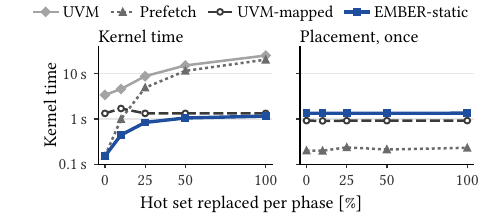}
  \vspace{-0.55cm}
  \caption{Runtime with a hot set that moves between phases (GH200, $14$\,GiB window of $142$\,GiB, eight phases).}
  \label{fig:sweep}
\end{figure}

\section{Evaluation}
\label{sec:eval}

We evaluate \sys along five dimensions: its hybrid address space benefits (\autoref{sec:eval-micro}), its performance on three distinct use-cases (\autoref{sec:eval-apps}--\autoref{sec:eval-db}), its behavior across GPUs and interconnects (\autoref{sec:eval-gpus}), the value of runtime decisions against profiled static ones (\autoref{sec:eval-ablation}), and the trade-off between its two variants (\autoref{sec:eval-overhead}).

\subsection{Experimental Setup}
\label{sec:setup}

\myparagraph{Testbed}
We conduct our experiments on four machines with different GPUs that exemplify three deployment setups (\autoref{tab:hw}): a consumer card, two datacenter GPUs, and a superchip. %
Their device-memory-to-interconnect bandwidth ratio ranges from ${\sim}9\times$ to ${\sim}150\times$.
We set the working set sizes to multiples of GPU memory; for GPUs with more memory, we reserve the surplus (\texttt{cudaMalloc}) before our runs.

\myparagraph{Variants}
\autoref{tab:variants} lists the four baselines available on stock CUDA (\autoref{sec:bg:control}) and the two variants of \sys.
We additionally compare against HELM~\cite{DBLP:conf/sc/JonesAG25}, which profiles an application and selects a placement policy for each allocation.

\myparagraph{Workloads}
We exercise \sys in a variety of workloads: six \textit{data-intensive} applications (from UVM-Bench~\cite{DBLP:journals/corr/abs-2007-09822}, radix sort from Thrust, BFS from EMOGI~\cite{DBLP:journals/pvldb/MinMQXEH20}, and our bitonic sort), \textit{LLM inference} with \texttt{llama-bench}, and \textit{database analytics} running SSB queries on Crystal~\cite{DBLP:conf/sigmod/ShanbhagMY20}.
We also run HELM~\cite{DBLP:conf/sc/JonesAG25} on five of its seven benchmarks (\textit{stream} and \textit{fdtd-2d} run out of memory) and a microbenchmark with a varying hot set.

\myparagraph{Metrics}
We measure total application runtime from a cold allocation.
Benchmark runs are capped at $1800$\,s, and a speedup over a capped run is a lower bound, marked with \dag.

\subsection{Hybrid Address Space}
\label{sec:eval-micro}

First, we evaluate the effect of \sys's hybrid address space.

\myparagraph{Methodology}
On GH200, a kernel reads a $14$\,GiB window of a $142$\,GiB managed buffer in eight phases, once per phase. After each one, the window slides forward by a fixed share of its size, from $0\%$, where the hot set never moves, to $100\%$, where it is replaced between phases.
We compare the static variants to isolate the effect of the hybrid address space.

\myparagraph{Results}
\autoref{fig:sweep} shows the results.
The Prefetch variant is fastest when the hot set does not change ($0.15$\,s), but slows down considerably once it shifts, since each prefetch is stale by the next phase.
UVM-mapped's runtime remains stable ($\approx1.33$\,s), incurring the interconnect cost on every access. 
Stock UVM, regardless of the data shift, remains the slowest.
\sys-static matches the Prefetch when the hot set is static and degrades at full drift, but still outperforms all baselines.
Its advice and prefetch add $1.33$\,s once (\autoref{fig:sweep}, right), a cost that gets amortized over time via repeated reads.

\takeaway{\sys combines the benefits of prefetching, when the hot set does not change, with the robustness of mapping once it moves, at only a one-time setup cost.}

\subsection{Use Case \#1: Data-Intensive Applications}
\label{sec:eval-apps}
We examine how \sys behaves in data-intensive applications with large working sets and diverse access patterns.

\myparagraph{Methodology}
We run the six applications on the RTX~5080, the A40, and the H200 at $0.5\times$ and ${\sim}1.5\times$ GPU memory, and report speedups over stock UVM; absolute runtimes are reported in \autoref{app:abs}.
The applications are grouped by access pattern, sequential (pathfinder, k-means, CNN), strided (bitonic sort), and random (radix sort, BFS).

\begin{figure*}[t]
  \centering
  \includegraphics[width=\textwidth]{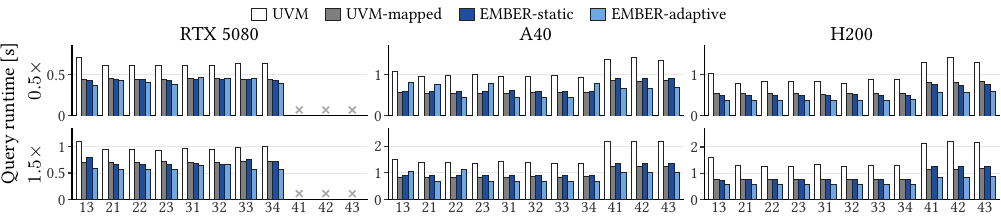}
  \vspace{-0.4cm}
  \caption{Runtime of SSB queries~\cite{DBLP:conf/tpctc/ONeilOCR09} on Crystal. $\times$: query exceeds the RTX~5080 host's memory.}
  \vspace{-0.2cm}
  \label{fig:db}
\end{figure*}

\begin{figure}[t]
  \centering
  \includegraphics[width=\columnwidth]{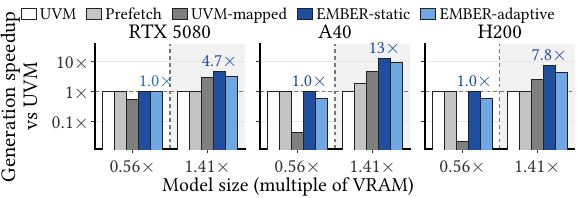}
  \vspace{-0.55cm}
  \caption{LLM token generation at $0.56\times$ and $1.41\times$ device memory, speedup over UVM (log scale), with every GPU capped to $16$\,GB. Labels show the faster variant.}
  \label{fig:llm}
\end{figure}

\myparagraph{Results}
\autoref{fig:bar} shows the results for normal and oversubscribed scenarios.
While the data fits, \sys-static achieves a mean speedup of $1.9$--$3.0\times$ over stock UVM, with Prefetch at $1.3$--$2.3\times$; UVM-mapped remains below stock UVM at $0.3$--$0.5\times$.
\sys-adaptive reaches $1.6$--$1.7\times$, as it starts mapped and promotes only based on measured demand.

Under oversubscription, \sys-static achieves a mean speedup over stock UVM of $9.12\times$ on the RTX~5080, $6.83\times$ on the A40, and $10.62\times$ on the H200, and \sys-adaptive $8.83\times$, $5.52\times$, and $8.01\times$.
UVM-mapped reaches $3.1$--$4.7\times$ and remains $2.2$--$2.5\times$ behind \sys-static, since every access crosses the interconnect.
Prefetch falls below UVM on the RTX~5080 ($0.95\times$), as each prefetch evicts data that is needed again once the footprint exceeds GPU memory.
\sys's performance improvement grows with the irregularity of the access pattern; from $1.8$--$10\times$ on the sequential applications to $11$--$225\times$ on the random ones, since UVM's prefetcher (\autoref{sec:bg:tbn}) performs well on sequential access but mostly migrates unread pages on random access.
Further, applications with sequential and strided access patterns often re-read their data resulting in an up to $10\times$ speedup for \sys. %

\takeaway{On data-intensive applications, \sys's variants outperform stock UVM on every GPU class, regardless of whether the data fits or not.}

\subsection{Use Case \#2: LLM Inference}
\label{sec:eval-llm}

Next, we verify whether \sys enables an LLM inference workload to function even under oversubscription.

\myparagraph{Methodology}
On the three discrete GPUs, \texttt{llama-bench} runs chatbot-style generation, a $512$-token prompt followed by $128$ generated tokens, for two models, Qwen2.5~14B and Qwen3~32B.
We cap every GPU at $16$\,GB, so the same model occupies the same footprint on each, and we report token generation alone, since it re-reads every weight per token, whereas prefill reads them once and behaves like a scan.
Explicit staging cannot load a model beyond GPU memory.%

\myparagraph{Results}
While the model fits (\autoref{fig:llm}), \sys-static performs similarly to UVM and prefetching (within $2\%$), while mapping the weights is $2\times$ slower on the RTX~5080 and $23$--$48\times$ on the A40 and H200, due to the slower interconnect.
In the case of GPU oversubscription ($1.41\times$), stock UVM and prefetching fall to $0.2$--$0.5$ tokens/s, turning a $128$-token reply into minutes. In contrast, \sys-static allows for a $4.7\times$ faster token generation on the RTX~5080 and $8$--$13\times$ on the A40 and H200. \sys-adaptive's respective rates are $3$--$10\times$.
UVM-mapped bounces back ($2.5$--$4.8\times$ over UVM), but still below \sys, as expected for a strategy bounded by the interconnect.
No strategy exceeds $3$ tokens/s, as \sys is bounded by the interconnect and stock UVM by page faults.

\takeaway{\sys executes inference on large models $3$-$13\times$ faster than UVM and up to $3\times$ than UVM-mapped, at footprints where explicit staging cannot load the model.}

\subsection{Use Case \#3: Database Analytics}
\label{sec:eval-db}
We evaluate \sys on analytical queries whose hot sets vary over time to quantify its adaptability benefits.

\myparagraph{Methodology}
We run the SSB queries on Crystal on the RTX~5080, the A40, and the H200, at $0.5\times$ and $1.5\times$ GPU memory; q11 and q12 emitted no runtime summary.
For a fair comparison, we cap the A40 and the H200 to the RTX~5080's $16$\,GiB by manually reserving the remaining capacity.

\myparagraph{Results}
As observed in \autoref{fig:db}, all baselines outperform stock UVM on every query and GPU, by $1.2$--$2.8\times$. 
\sys-adaptive is the fastest in 51 of the 60 measured query--GPU--size combinations, and in all 22 on the H200.
Its mean speedup over UVM is $1.50$--$1.66\times$ on the RTX~5080, $1.75$--$1.92\times$ on the A40, and $2.26$--$2.36\times$ on the H200, where the respective values for UVM-mapped are at most $1.40$, $1.72$, and $1.75\times$.
\sys-static stays within about $10\%$ of the performance of UVM-mapped.
This difference is attributable to the interconnect.
Precisely, the slower it is relative to GPU memory, the more a promoted column can save (\autoref{tab:hw}).
A single scan reads each column once, so \sys-static preserves the loaded hot set based on the first allocation, and does not adjust to the application's requirements. 
\sys-adaptive instead promotes the columns a query actually reads.

\takeaway{On analytical queries, \sys-adaptive proves to be the fastest; it has $1.50$--$2.36\times$ speedup over UVM, since it promotes only the columns a query needs.}

\begin{figure}[t]
  \centering
  \includegraphics[width=\columnwidth]{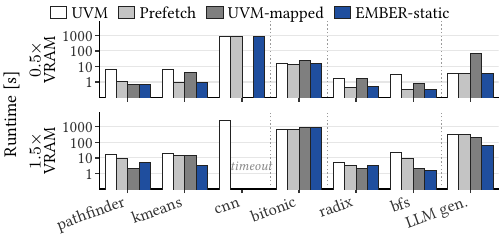}
  \vspace{-0.55cm}
  \caption{Runtime on the GH200 (log scale).}
  \label{fig:gpus}
\end{figure}

\subsection{GPU Classes \& Coherent Interconnects}
\label{sec:eval-gpus}

We execute applications on the GH200 to determine if a faster, coherent interconnect eliminates the benefits of \sys.

\myparagraph{Methodology}
We repeat the six applications and LLM generation on the GH200 at $0.5\times$ and $1.5\times$ GPU memory, reporting absolute runtime, and for the LLM the time to generate $128$ tokens.
We report \sys-static alone, since we had no bare-metal access and could not run \sys-adaptive (\autoref{sec:eval-limits}).

\myparagraph{Results}
\autoref{fig:gpus} presents the results.
UVM falls behind the baselines in almost all cases, but does not collapse past the VRAM line: \sys-static's geometric-mean gain over it holds near $3.5\times$ at both footprints, where on PCIe it reaches $6.8$--$10.6\times$.
Over a coherent link, the GPU reads a non-resident page in place (\autoref{sec:bg:coherent}), so the fault-and-evict loop largely disappears.
UVM-mapped remains costly when data is reused (i.e., $4.7\times$ slower than \sys on k-means, $21\times$ on LLM generation).
Past the VRAM capacity, UVM-mapped beats \sys on the workloads that read data once (pathfinder, radix sort), since a single pass never amortizes the migration.
However, on k-means and LLM generation, which read the same data multiple times, \sys executes in $3.1$\,s and generates the $128$ tokens in $62$\,s, compared to $14.2$\,s and $211$\,s for UVM-mapped. 
This performance gain stems from the data placement and reuse frequency, as, according to our measurements, the workloads, even on the mapped case, do not saturate the interconnect bandwidth (i.e., $10$\,GiB/s on k-means).

\takeaway{A fast, coherent link makes reading from host memory competitive for data that is read once, but reused data still benefits from GPU memory, highlighting the need for \sys's data placement strategies.}

\begin{figure}[t]
  \centering
  \includegraphics[width=\columnwidth]{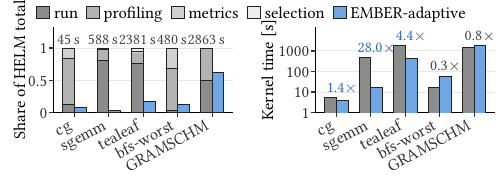}
  \vspace{-0.7cm}
  \caption{Comparison of HELM and \sys (RTX~5080).}
  \label{fig:helm}
\end{figure}

\subsection{Comparison with Profiled Placement}
\label{sec:eval-ablation}
Next, we evaluate \sys-adaptive to quantify the benefits of its dynamic data placement and reclamation policy.

\myparagraph{Methodology}
We compare \sys with HELM~\cite{DBLP:conf/sc/JonesAG25}, which profiles an application and assigns each allocation a policy (e.g., \textit{migrate}, \textit{pin to host}, \textit{pin to device}).
We measure both the profiling time to decide on data placement and the actual kernel time of the application.
We also tried ARIADNE~\cite{DBLP:conf/hpca/ShinBPK26}, the closest in-driver system, but could not run it, since it builds against driver release $535.86$ and relies on legacy interfaces.

\myparagraph{Results}
\sys-adaptive outperforms HELM on all five benchmarks (\autoref{fig:helm}, left), by a mean of $7.9\times$, even though HELM profiles each application first.
Profiling, metric processing, and selection take $39$ to $1449$\,s, account for $18\%$ to $95\%$ of HELM's end-to-end runtime, and heavily depend on the problem size.
The choice is also unstable, as repeating the profiling run can tag the same allocation differently. %
Removing that cost and comparing kernel time alone (\autoref{fig:helm}, right), \sys-adaptive remains faster on three of five benchmarks, by up to $28.0\times$ with a mean of $2.1\times$.
It is slower on bfs and gramschm, where HELM's choice fits the workload. %
For the rest, \sys's benefits mainly come from the placement granularity, since HELM's policy-per-allocation cannot adjust when hot and cold data are interleaved within one.

\takeaway{\sys's adaptive data placement matches or exceeds profiled per-allocation policies without a profiling run and without application changes.}

\subsection{\sys-adaptive vs \sys-static}
\label{sec:eval-overhead}
\begin{figure}[t]
  \centering
  \includegraphics[width=\columnwidth]{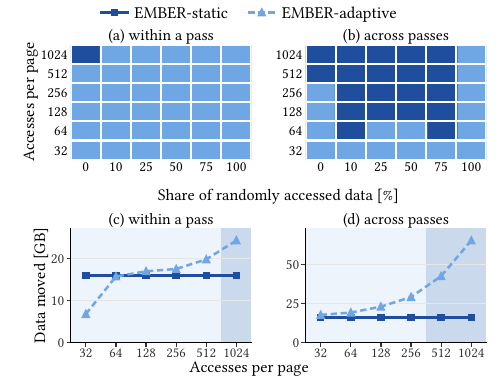}
  \vspace{-0.55cm}
  \caption{Dominant variant (a,~b) and data moved at $0\%$ randomness (c,~d), by reuse per page (RTX~5080, $1.3\times$ GPU memory). Shading marks where \sys-static is faster.}
  \label{fig:winner}
\end{figure}

Finally, we aim to identify when each \sys variant is the right choice for an application.

\myparagraph{Methodology}
On the RTX~5080, a kernel reads a $20$\,GB managed buffer, at $1.3\times$ GPU memory, with a configurable share accessed in random order and the rest in a linear sweep, and each page accessed $32$ to $1024$ times.
We sweep two reuse patterns.
In the first, all accesses to a page occur in a single pass.
In the second, each pass accesses every page once, and reuse comes from repeated passes. We report runtime over the whole grid and, at $0\%$ randomness, the data movement.

\myparagraph{Results}
\sys-static incurs a one-time $0.74$\,s for advising the buffer and prefetching a $16$\,GB prefix, whereas \sys-adaptive incurs one migration per promotion. %
\sys-adaptive is faster at low reuse ($32$ accesses), by $1.3$--$4\times$, where the prefetch is not amortized, and at both ends of the randomness axis, where a pure sweep is a sliding window, tracked by \sys's counters, while a purely random pattern implies that there is no hot set worth pinning.
\sys-static is faster in between, by up to $3.2\times$ once a page is reused $128$ times or more, due to its pinned hot set. %
Static pays its overhead once and adaptive on every promotion, so which variant wins follows reuse (\autoref{fig:winner}a,b).
In terms of data movement (\autoref{fig:winner}c,d), \sys-static migrates its $16$\,GB prefix once, while \sys-adaptive moves $6.8$ to $24.4$\,GB inside a pass and $17.7$ to $65.2$\,GB when accesses span different passes, since every pass demotes regions that the next one promotes again.

Overall, \emph{when the data fits} in GPU memory, \sys-adaptive runs $1.2$--$1.8\times$ slower than \sys-static. However, \sys mainly targets oversubscribed workloads, where the hot set typically exceeds GPU memory and must be adjusted at runtime, thereby favoring \sys's adaptive variant.
On \emph{integration cost}, \sys-adaptive requires \emph{no} application changes, only a module load, while \sys-static adds $8$--$28$ lines per application, under $5\%$ in every case, none of them in a GPU kernel, in launch code, or in an algorithm (\autoref{app:loc}).

\takeaway{\sys-static is the better choice when a stable hot set is frequently re-read, and \sys-adaptive otherwise, when reuse is low, confined to a single pass, or spread too uniformly to leave a hot set worth pinning.}

\section{Limitations and Future Work}
\label{sec:eval-limits}
\myparagraph{Library buffers}
\sys places only memory allocated via \texttt{cudaMallocManaged}.
Buffers a library allocates internally, such as Thrust's temporaries in radix sort, take device memory that neither variant can see or place.
\sys-static suffers more, since it sizes its prefix from one free-memory query before launch, which a later library allocation invalidates.
\sys-adaptive sizes its headroom at runtime, so it stops promoting when a library takes the memory (\autoref{sec:design:adaptive}).

\myparagraph{Device-access observability}
\sys cannot observe that a resident region has gone cold; it infers this from the demand the region showed before promotion, decayed over time (\autoref{sec:design:adaptive}).
A region that is still hot can thus be reclaimed by mistake and moved twice.
The open driver contains an undocumented \emph{vidmem access-bit buffer} that reports access and dirty bits over device memory~\cite{nvidia_open_gpu_kernel_modules}, reachable only from kernel space and limited to eight address ranges at a time.
It is disabled by default on Blackwell, where it is reported to cause undefined behaviour on some parts, so we do not use it.
Feeding it into reclamation would replace inference with measurement for those ranges; this and other policy refinements are orthogonal to \sys and left for future work.

\myparagraph{Deployment}
\sys-adaptive needs bare-metal access to load a kernel module.
Our GH200 instance is a cloud VM without that access, which is why \autoref{sec:eval-gpus} reports \sys-static alone.
The module also builds against the internal interfaces of an open-source driver release and must be rebased for each new one.
The rebase is usually mechanical, but a release that changes those interfaces can make a mechanism harder to implement or unavailable, as it did for ARIADNE (\autoref{sec:eval-ablation}).
Nothing in the design is specific to CUDA: the hybrid address space needs only driver-controlled migration and host mapping, which other stacks also provide (e.g., AMD ROCm \cite{amd_rocm}).
\sys-adaptive additionally needs a non-faulting signal of remote access frequency.

\myparagraph{Multi-GPU}
\sys places data between one GPU and host memory.
On a multi-GPU node, peer memory forms a third tier between the two, so a region has three homes rather than two, and each notification has to be attributed to the GPU that raised it.
We aim to explore it as future work.

\section{Related Work}
\label{sec:related}

\myparagraph{UVM optimizations}
Prior work analyzes the cost of UVM's faults, prefetching, and eviction~\cite{DBLP:conf/hpec/LandaverdeZCH14, DBLP:conf/sc/MishraLKFC17, DBLP:conf/ispass/VeselyBOLB16, DBLP:conf/ipps/Allen021, DBLP:conf/sc/Allen021, DBLP:conf/isca/GangulyZYM19, DBLP:conf/sc/ChienPM19, DBLP:journals/tjs/KnapC19} and addresses its weaknesses with hardware changes~\cite{DBLP:conf/hpca/ZhengNZSK16, DBLP:conf/micro/Ausavarungnirun17, DBLP:journals/taco/ZiabariSMSAUKJK16, DBLP:journals/corr/abs-2411-05309} as well as eviction and throttling policies~\cite{DBLP:conf/asplos/0015ARZMGY19, DBLP:journals/tcad/YuCHQW20}.
Adaptive runtimes retune prefetching and eviction behavior~\cite{DBLP:conf/date/GangulyMY21, DBLP:conf/ispass/GoLKLYR23, DBLP:conf/ipps/0003CHQ0W20, DBLP:conf/asplos/KimSGHK20, DBLP:conf/iiswc/ZhuCVHCR22, DBLP:journals/cal/ShinBPK25}, and other works decide on data migration based on heuristics, such as hardware counters~\cite{DBLP:conf/isca/LinFCJ25}, fault history~\cite{DBLP:conf/asplos/JungKL23}, or profiling~\cite{lin_autouvm_2026}.
All of them keep accesses on the migration path, preserving the reclamation traffic. In contrast, \sys leaves the cold set mapped in host memory, removing that traffic while adjusting the GPU-resident set.

\myparagraph{Data movement and placement}
Several systems combine mapped access with migration. %
Some take data placement decisions ahead of time based on compile-time analysis or offline profiling runs~\cite{DBLP:conf/micro/BCVB24, DBLP:conf/sc/JonesAG25}.
Others decide at runtime based on page access count threshold~\cite{DBLP:conf/ipps/GangulyZYM20}, compiler instrumentation that identifies sparsely used pages~\cite{DBLP:conf/systor/ChangKS21}, or the number of GPU cores faulting on a block concurrently~\cite{DBLP:journals/cal/ShinBPK25, DBLP:conf/hpca/ShinBPK26}.
Another work aims to restore the visibility of GPU counters by keeping one page of each region mapped in host memory, at the cost of one sampled page per region, and makes placement decisions based on those~\cite{DBLP:conf/isca/BSB26}.
\sys decides at runtime and chooses mapping as the default, with regions migrated only once the demand based on the GPU counters instructs it, requiring no offline profiling, additional faults, or sampling.

\myparagraph{Application-level placement}
Other systems encode data placement within the application, for out-of-memory graph processing~\cite{DBLP:journals/pvldb/MinMQXEH20, DBLP:conf/eurosys/Sabet0G20, DBLP:journals/pvldb/GeraKSKB20}, DNN training, and LLM inference, with a dataflow known in advance~\cite{DBLP:conf/hipc/AwanCSLP18, DBLP:conf/micro/RhuGCZK16, DBLP:conf/sc/RajbhandariRRSH21, DBLP:conf/icml/0007ZYLRCLRSZ23}.
\sys keeps the application intact and requires no such knowledge.
Multi-GPU placement and memory pooling~\cite{DBLP:conf/hpca/WangLJYT24, DBLP:conf/hpca/WangLZE0JT25, DBLP:conf/usenix/ChoiKJAJKA22, DBLP:conf/ipps/LavoreMADBS24} is orthogonal, as the GPUs together can fit the working set.

\myparagraph{Tiered memory}
CPU tiering classifies pages from hint faults or sampled events and migrates them between DRAM and a slower memory~\cite{DBLP:conf/sosp/RaybuckSZEP21, DBLP:conf/sosp/LeeMME23, DBLP:conf/eurosys/QiZHH0H025, DBLP:conf/eurosys/0015XRSKL24}. TPP demotes proactively to keep headroom~\cite{DBLP:conf/asplos/MarufWDWABPCKC23}, an idea \sys adopts.
These designs cannot be directly applied to GPU tiering, as the GPU's ${\sim}256$ counters observe only host memory, and its slow tier sits $15$--$150\times$ below GPU memory.
Therefore, \sys aggregates counter notifications into per-region demand that decays over time, and infers when resident data turns cold.

\section{Conclusion}
\label{sec:conclusion}

UVM enables applications to oversubscribe GPU memory, but performance collapses once the working set exceeds the GPU's memory capacity.
Its placement decisions are reactive (i.e., per-fault) and do not consider access patterns, leading to increased reclamation traffic under memory pressure and execution pipeline stalls.
To tackle this, we present \sys, a GPU memory management framework tailored for oversubscribed workloads, that preserves the UVM memory and programming model.
\sys splits allocations into smaller regions and initially maps them into host memory, which is directly reachable by the GPU, migrating a region only after the GPU hardware counters report it as hot.
\sys adjusts its placement decisions at runtime and proactively reclaims memory, keeping eviction off the critical path.
Across various workload classes, GPUs, and interconnects, \sys outperforms existing UVM placement strategies under oversubscription as well as optimized profile-guided frameworks without requiring a profiling run.

\myparagraph{Artifact and appendix} \sys{} will be publicly available. Its artifact covers detailed experimental results (\autoref{app:abs}) and highlights \sys's integration costs (\autoref{app:loc}). %

\begin{acks}
\noindent\raisebox{-3pt}{\includegraphics[height=12pt,trim=37 72 31 75,clip]{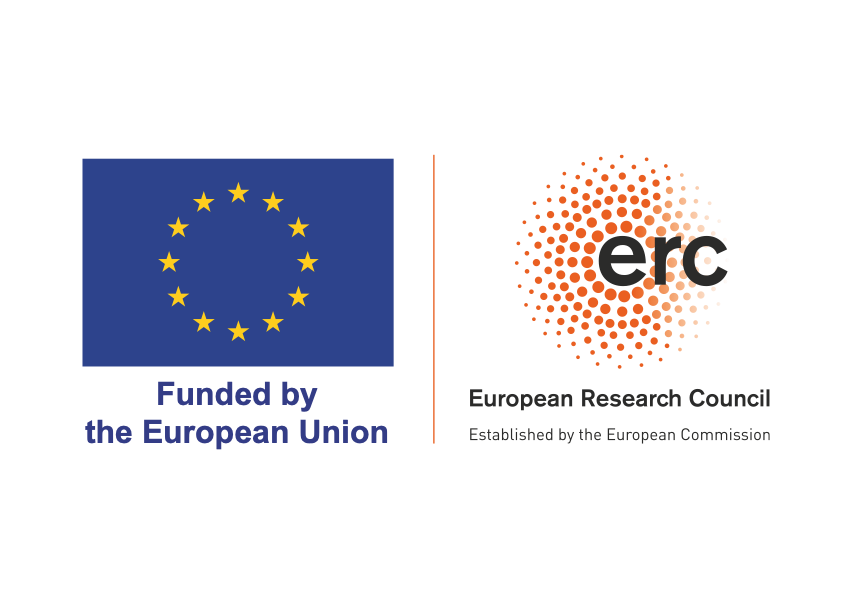}}
Funded by the European Union (ERC, FDS, 101164556). Views and opinions expressed are however those of the authors only and do not necessarily reflect those of the European Union or the European Research Council Executive Agency. Neither the European Union nor the granting authority can be held responsible for them.
\end{acks}

\balance
\bibliographystyle{ACM-Reference-Format}
\bibliography{references}

\ifthenelse{\boolean{appendix}}
{
\clearpage
\pagestyle{empty}
\appendix
\onecolumn
\section{Detailed Experimental Results}
\label{app:abs}
The tables in this appendix give the absolute numbers behind every figure in \S\ref{sec:eval}, generated from the same measurement files the figures are drawn from.
Runtimes are in seconds unless otherwise stated; $>$1800 marks a run that hit its cap, and a dash marks a combination that was not run.
\autoref{tab:abs-apps} covers the applications on the PCIe GPUs (\autoref{fig:bar}), \autoref{tab:abs-gh200} the GH200 (\autoref{fig:gpus}), \autoref{tab:abs-llm} LLM generation (\autoref{fig:llm}), \autoref{tab:abs-db-half} and \autoref{tab:abs-db-full} the SSB queries (\autoref{fig:db}), \autoref{tab:abs-helm} the HELM comparison (\autoref{fig:helm}), and \autoref{tab:abs-sweep} the moving-hot-set microbenchmark (\autoref{fig:sweep}).

\begin{table}[H]
  \centering
  \caption{Application runtime in seconds on the three PCIe GPUs (\autoref{fig:bar}). The $0.5\times$ inputs are $8$\,GiB; the ${\sim}1.5\times$ inputs are $22$\,GiB (RTX~5080), $24$\,GiB (A40), and $16$\,GiB (H200) at $16$\,GiB device memory. Columns per GPU: UVM, Prefetch, UVM-mapped, \sys-static, \sys-adaptive. $>$1800: hit the cap.}
  \label{tab:abs-apps}
  \scriptsize
  \setlength{\tabcolsep}{3pt}
  \begin{tabular}{@{}llrrrrrrrrrrrrrrr@{}}
    \toprule
    Application & Footprint & \multicolumn{5}{c}{RTX~5080} & \multicolumn{5}{c}{A40} & \multicolumn{5}{c}{H200} \\
    \cmidrule(lr){3-7}\cmidrule(lr){8-12}\cmidrule(lr){13-17}
     &  & UVM & Prefetch & UVM-mapped & static & adaptive & UVM & Prefetch & UVM-mapped & static & adaptive & UVM & Prefetch & UVM-mapped & static & adaptive \\
    \midrule
    pathfinder & $0.5\times$ & 0.52 & 0.64 & 0.99 & 0.56 & 0.43 & 0.84 & 0.38 & 0.43 & 0.38 & 0.44 & 0.95 & 0.34 & 0.42 & 0.33 & 0.44 \\
    k-means & $0.5\times$ & 0.62 & 0.38 & 2.72 & 0.38 & 0.45 & 1.04 & 0.49 & 3.87 & 0.49 & 0.63 & 1.13 & 0.54 & 2.95 & 0.60 & 0.52 \\
    cnn & $0.5\times$ & 596 & 557 & $>$1800 & 84.8 & 64.9 & 671 & 240 & $>$1800 & 120 & 123 & 731 & 793 & $>$1800 & 142 & 254 \\
    bitonic sort & $0.5\times$ & 4.78 & 4.84 & 69.0 & 4.97 & 4.64 & 6.87 & 6.67 & 72.5 & 6.97 & 9.11 & 2.72 & 2.46 & 71.3 & 2.68 & 3.60 \\
    radix sort & $0.5\times$ & 0.68 & 0.62 & 2.40 & 0.58 & 0.74 & 1.33 & 1.09 & 4.71 & 1.08 & 1.64 & 1.29 & 0.85 & 2.53 & 0.79 & 1.22 \\
    bfs & $0.5\times$ & 0.56 & 0.15 & 0.39 & 0.15 & 0.44 & 0.95 & 0.21 & 0.41 & 0.21 & 0.58 & 0.85 & 0.05 & 0.40 & 0.05 & 0.43 \\
    \midrule
    pathfinder & ${\sim}1.5\times$ & 2.85 & 4.88 & 2.80 & 2.38 & 1.61 & 2.69 & 2.83 & 1.28 & 1.20 & 1.30 & 2.27 & 2.16 & 0.84 & 0.76 & 0.91 \\
    k-means & ${\sim}1.5\times$ & 4.34 & 3.22 & 7.59 & 1.13 & 1.29 & 6.03 & 6.94 & 13.3 & 2.28 & 2.88 & 4.36 & 5.72 & 7.18 & 1.39 & 1.20 \\
    cnn & ${\sim}1.5\times$ & $>$1800 & $>$1800 & $>$1800 & 234 & 177 & $>$1800 & $>$1800 & $>$1800 & 357 & 350 & $>$1800 & $>$1800 & $>$1800 & 282 & 492 \\
    bitonic sort & ${\sim}1.5\times$ & 552 & 553 & 209 & 75.6 & 99.9 & 715 & 718 & 239 & 106 & 134 & 550 & 546 & 157 & 84.6 & 110 \\
    radix sort & ${\sim}1.5\times$ & $>$1800 & $>$1800 & 7.98 & 9.11 & 8.01 & 487 & 504 & 14.3 & 11.9 & 15.3 & 1150 & 1124 & 5.32 & 5.09 & 5.87 \\
    bfs & ${\sim}1.5\times$ & 8.15 & 8.61 & 1.08 & 0.72 & 1.29 & 10.9 & 11.0 & 1.24 & 0.89 & 1.46 & 10.2 & 10.4 & 1.21 & 0.62 & 1.27 \\
    \bottomrule
  \end{tabular}
\end{table}

\begin{table}[H]
  \centering
  \caption{GH200 runtime in seconds (\autoref{fig:gpus}): applications at $48$ and $142$\,GiB, and generating $128$ tokens with models at $0.41\times$ and $1.37\times$ of the $96$\,GiB. cnn was capped at $3600$\,s.}
  \label{tab:abs-gh200}
  \scriptsize
  \setlength{\tabcolsep}{3pt}
  \begin{tabular}{@{}llrrrr@{}}
    \toprule
    Workload & Footprint & UVM & Prefetch & UVM-mapped & \sys-static \\
    \midrule
    pathfinder & $0.5\times$ & 6.38 & 1.04 & 0.75 & 0.72 \\
    k-means & $0.5\times$ & 6.44 & 0.89 & 4.15 & 0.88 \\
    cnn & $0.5\times$ & 874 & 847 & $>$3600 & 805 \\
    bitonic sort & $0.5\times$ & 15.7 & 14.3 & 23.3 & 14.7 \\
    radix sort & $0.5\times$ & 1.66 & 0.45 & 1.71 & 0.50 \\
    bfs & $0.5\times$ & 3.30 & 0.33 & 0.76 & 0.33 \\
    LLM, 128 tokens & $0.5\times$ & 3.49 & 3.49 & 74.2 & 3.50 \\
    \midrule
    pathfinder & $1.5\times$ & 17.8 & 9.03 & 2.21 & 4.82 \\
    k-means & $1.5\times$ & 19.7 & 14.1 & 14.2 & 3.13 \\
    cnn & $1.5\times$ & 2594 & $>$1800 & $>$3600 & $>$3600 \\
    bitonic sort & $1.5\times$ & 717 & 709 & 861 & 868 \\
    radix sort & $1.5\times$ & 5.17 & 3.38 & 2.04 & 3.31 \\
    bfs & $1.5\times$ & 23.1 & 9.74 & 2.21 & 1.49 \\
    LLM, 128 tokens & $1.5\times$ & 328 & 327 & 211 & 61.4 \\
    \bottomrule
  \end{tabular}
\end{table}

\begin{table}[H]
  \centering
  \caption{LLM token generation in tokens per second (\autoref{fig:llm}), $128$ tokens after a $512$-token prompt, every GPU capped to $16$\,GB.}
  \label{tab:abs-llm}
  \scriptsize
  \setlength{\tabcolsep}{3pt}
  \begin{tabular}{@{}lllrrrrr@{}}
    \toprule
    GPU & Size & Model & UVM & Prefetch & UVM-mapped & \sys-static & \sys-adaptive \\
    \midrule
    RTX~5080 & 0.56$\times$ & \texttt{Qwen2.5-14B-Q4\_K\_M} & 84.8 & 84.8 & 45.2 & 84.8 & 85.0 \\
    RTX~5080 & 1.41$\times$ & \texttt{Qwen3-32B-Q5\_K\_S} & 0.45 & 0.45 & 1.37 & 2.10 & 1.46 \\
    \midrule
    A40 & 0.56$\times$ & \texttt{Qwen2.5-14B-Q4\_K\_M} & 54.7 & 54.7 & 2.41 & 55.0 & 32.7 \\
    A40 & 1.41$\times$ & \texttt{Qwen3-32B-Q5\_K\_S} & 0.21 & 0.40 & 1.00 & 2.71 & 2.07 \\
    \midrule
    H200 & 0.56$\times$ & \texttt{Qwen2.5-14B-Q4\_K\_M} & 113 & 113 & 2.36 & 113 & 66.5 \\
    H200 & 1.41$\times$ & \texttt{Qwen3-32B-Q5\_K\_S} & 0.38 & 0.38 & 0.96 & 2.94 & 1.73 \\
    \bottomrule
  \end{tabular}
\end{table}

\begin{table}[H]
  \centering
  \caption{SSB query runtime in seconds on Crystal at $0.5\times$ device memory (\autoref{fig:db}), every GPU capped to $16$\,GiB. Columns per GPU: UVM, UVM-mapped, \sys-static, \sys-adaptive. q41--q43 exceed the RTX~5080 host's memory.}
  \label{tab:abs-db-half}
  \scriptsize
  \setlength{\tabcolsep}{3pt}
  \begin{tabular}{@{}lrrrrrrrrrrrr@{}}
    \toprule
    Query & \multicolumn{4}{c}{RTX~5080} & \multicolumn{4}{c}{A40} & \multicolumn{4}{c}{H200} \\
    \cmidrule(lr){2-5}\cmidrule(lr){6-9}\cmidrule(lr){10-13}
     & UVM & UVM-mapped & static & adaptive & UVM & UVM-mapped & static & adaptive & UVM & UVM-mapped & static & adaptive \\
    \midrule
    13 & 0.71 & 0.45 & 0.43 & 0.38 & 1.08 & 0.57 & 0.59 & 0.80 & 1.05 & 0.54 & 0.50 & 0.38 \\
    21 & 0.61 & 0.45 & 0.44 & 0.43 & 0.96 & 0.56 & 0.60 & 0.77 & 0.79 & 0.55 & 0.50 & 0.38 \\
    22 & 0.61 & 0.45 & 0.44 & 0.40 & 0.97 & 0.56 & 0.60 & 0.45 & 0.85 & 0.55 & 0.50 & 0.38 \\
    23 & 0.61 & 0.45 & 0.44 & 0.38 & 1.00 & 0.55 & 0.60 & 0.79 & 0.84 & 0.54 & 0.51 & 0.38 \\
    31 & 0.61 & 0.45 & 0.44 & 0.46 & 0.96 & 0.56 & 0.61 & 0.45 & 0.83 & 0.54 & 0.51 & 0.38 \\
    32 & 0.61 & 0.45 & 0.44 & 0.46 & 0.96 & 0.57 & 0.60 & 0.46 & 0.79 & 0.54 & 0.52 & 0.39 \\
    33 & 0.63 & 0.45 & 0.44 & 0.46 & 0.98 & 0.56 & 0.60 & 0.46 & 0.88 & 0.55 & 0.50 & 0.39 \\
    34 & 0.63 & 0.45 & 0.43 & 0.40 & 0.94 & 0.56 & 0.59 & 0.79 & 0.89 & 0.55 & 0.51 & 0.39 \\
    41 & -- & -- & -- & -- & 1.38 & 0.85 & 0.90 & 0.67 & 1.30 & 0.82 & 0.76 & 0.57 \\
    42 & -- & -- & -- & -- & 1.42 & 0.84 & 0.91 & 0.67 & 1.43 & 0.81 & 0.74 & 0.57 \\
    43 & -- & -- & -- & -- & 1.36 & 0.86 & 0.91 & 0.68 & 1.31 & 0.84 & 0.77 & 0.59 \\
    \bottomrule
  \end{tabular}
\end{table}

\begin{table}[H]
  \centering
  \caption{SSB query runtime in seconds on Crystal at $1.5\times$ device memory (\autoref{fig:db}), every GPU capped to $16$\,GiB. Columns per GPU: UVM, UVM-mapped, \sys-static, \sys-adaptive. q41--q43 exceed the RTX~5080 host's memory.}
  \label{tab:abs-db-full}
  \scriptsize
  \setlength{\tabcolsep}{3pt}
  \begin{tabular}{@{}lrrrrrrrrrrrr@{}}
    \toprule
    Query & \multicolumn{4}{c}{RTX~5080} & \multicolumn{4}{c}{A40} & \multicolumn{4}{c}{H200} \\
    \cmidrule(lr){2-5}\cmidrule(lr){6-9}\cmidrule(lr){10-13}
     & UVM & UVM-mapped & static & adaptive & UVM & UVM-mapped & static & adaptive & UVM & UVM-mapped & static & adaptive \\
    \midrule
    13 & 1.11 & 0.71 & 0.79 & 0.59 & 1.50 & 0.84 & 0.89 & 1.06 & 1.61 & 0.78 & 0.75 & 0.57 \\
    21 & 0.95 & 0.71 & 0.67 & 0.57 & 1.40 & 0.84 & 0.90 & 0.67 & 1.28 & 0.78 & 0.76 & 0.57 \\
    22 & 0.96 & 0.70 & 0.67 & 0.57 & 1.41 & 0.85 & 0.91 & 1.13 & 1.25 & 0.78 & 0.76 & 0.57 \\
    23 & 0.94 & 0.72 & 0.67 & 0.57 & 1.37 & 0.83 & 0.90 & 0.67 & 1.26 & 0.78 & 0.76 & 0.57 \\
    31 & 0.97 & 0.70 & 0.68 & 0.64 & 1.42 & 0.83 & 0.91 & 0.67 & 1.32 & 0.77 & 0.77 & 0.57 \\
    32 & 0.95 & 0.70 & 0.67 & 0.67 & 1.38 & 0.85 & 0.89 & 0.68 & 1.28 & 0.78 & 0.77 & 0.58 \\
    33 & 0.99 & 0.72 & 0.75 & 0.57 & 1.35 & 0.85 & 0.90 & 0.68 & 1.30 & 0.78 & 0.76 & 0.58 \\
    34 & 1.01 & 0.72 & 0.71 & 0.57 & 1.35 & 0.85 & 0.90 & 0.68 & 1.31 & 0.78 & 0.76 & 0.58 \\
    41 & -- & -- & -- & -- & 2.17 & 1.26 & 1.34 & 1.00 & 2.13 & 1.16 & 1.25 & 0.86 \\
    42 & -- & -- & -- & -- & 2.20 & 1.25 & 1.34 & 1.00 & 2.23 & 1.17 & 1.25 & 0.86 \\
    43 & -- & -- & -- & -- & 2.18 & 1.25 & 1.34 & 1.02 & 2.19 & 1.18 & 1.28 & 0.88 \\
    \bottomrule
  \end{tabular}
\end{table}

\begin{table}[H]
  \centering
  \caption{HELM against \sys-adaptive on the RTX~5080 (\autoref{fig:helm}), seconds. Kernel: run time of the policy HELM selects; Total adds profiling, metrics, and selection. Single runs; \sys-adaptive is the mean of its runs.}
  \label{tab:abs-helm}
  \scriptsize
  \setlength{\tabcolsep}{3pt}
  \begin{tabular}{@{}lrrrrrr@{}}
    \toprule
    Benchmark & Profiling & Metrics & Selection & Kernel & Total & \sys-adaptive \\
    \midrule
    sgemm & 94.4 & 14.3 & 0.16 & 480 & 588 & 17.1 \\
    bfs-worst & 310 & 153 & 0.16 & 17.3 & 480 & 57.7 \\
    tealeaf & 444 & 2.26 & 136 & $>$1800 & $>$2381 & 413 \\
    cg & 32.2 & 6.60 & 0.23 & 5.44 & 44.5 & 3.78 \\
    GRAMSCHM & 1448 & 1.71 & 0.16 & 1413 & 2863 & 1800 \\
    \bottomrule
  \end{tabular}
\end{table}

\begin{table}[H]
  \centering
  \caption{Moving hot set on the GH200 (\autoref{fig:sweep}): kernel time over eight phases and one-time placement cost, seconds, by the share of the $14$\,GiB window replaced per phase.}
  \label{tab:abs-sweep}
  \scriptsize
  \setlength{\tabcolsep}{3pt}
  \begin{tabular}{@{}lrrrrrrr@{}}
    \toprule
    Drift [\%] & UVM & Prefetch & UVM-mapped & \sys-static & Pref.\ cost & Advise cost & \sys-static cost \\
    \midrule
    0 & 3.40 & 0.15 & 1.33 & 0.15 & 0.20 & 0.92 & 1.33 \\
    10 & 4.61 & 1.00 & 1.71 & 0.44 & 0.20 & 0.91 & 1.34 \\
    25 & 8.84 & 4.94 & 1.33 & 0.84 & 0.24 & 0.92 & 1.34 \\
    50 & 15.3 & 11.6 & 1.33 & 1.05 & 0.21 & 0.92 & 1.34 \\
    100 & 25.4 & 20.6 & 1.33 & 1.16 & 0.23 & 0.92 & 1.35 \\
    \bottomrule
  \end{tabular}
\end{table}

\section{Integration Cost}
\label{app:loc}
\autoref{tab:loc} lists the lines added to each application to run it under \sys-static (\S\ref{sec:eval-overhead}).

\begin{table}[H]
  \centering
  \footnotesize
  \caption{Integration cost of \sys-static.}
  \label{tab:loc}
  \begin{tabular}{lrrr}
    \toprule
    Application & Original LOC & Added & Increase \\
    \midrule
    pathfinder   & 358     & 12 & $<$4\% \\
    kmeans       & 423     & 12 & $<$3\% \\
    gramschm     & 274     & 13 & $<$5\% \\
    cnn          & 742     & 22 & $<$4\% \\
    bitonic-sort & 707     & 8  & $<$1.2\% \\
    radix-sort   & $>$3000 & 8  & $<$0.3\% \\
    bfs          & 657     & 28 & $<$5\% \\
    \bottomrule
  \end{tabular}
\end{table}

}
{

}

\end{document}